\documentclass[]{aa} 

\usepackage{graphicx}
\usepackage{txfonts}
\usepackage{natbib}
\usepackage{longtable}
\usepackage{lscape}

\usepackage{color}

\begin{document}
 
\title{Near-infrared extinction with discretised stellar colours}

\author{M. Juvela\inst{1} and J. Montillaud\inst{2}}

\institute{
Department of Physics, P.O.Box 64, FI-00014, University of Helsinki,
Finland, {\em mika.juvela@helsinki.fi}
\and
Institut UTINAM, CNRS UMR 6213, OSU THETA, Universit\'e de Franche-Comt\'e, 
41 bis avenue de l'Observatoire, 25000 Besan\c{c}on, France
}

\date{Received September 15, 1996; accepted March 16, 1997}

\abstract { 
Near-infrared (NIR) extinction remains one of the most reliable methods of measuring the
column density of dense interstellar clouds. Extinction can be estimated using the
reddening of the light of background stars. Several methods exist (e.g., NICE, NICER,
NICEST, GNICER) to combine observations of several NIR bands into extinction maps.
} 
{
We present a new method of estimating extinction based on NIR multiband observations and
examine its performance.
}
{
Our basic method uses a discretised version of the distribution of intrinsic
stellar colours directly. The extinction of individual stars and the average over a resolution
element are estimated with Markov chain Monte Carlo (MCMC) methods. Several variations
of the basic method are tested, and the results are compared to NICER calculations.
}
{
In idealised settings or when photometric errors are large, the results of the new
method are very close to those of NICER. Clear advantages can be seen when the
distribution of intrinsic colours cannot be described well with a single covariance
matrix. The MCMC framework makes it easy to consider additional effects such as those of
completeness limits and contamination by galaxies or foreground stars. \emph{A priori}
information about relative column density variations at sub-beam scales can result in a
significant increase in accuracy. For observations of high photometric precision,
the results could be further improved by considering the magnitude dependence of the
intrinsic colours.
}
{
The MCMC computations are time-consuming, but the calculation of large extinction maps is
already practical. The same methods can be used with direct optimisation, with
significantly less computational work. Faster methods, like NICER, perform very well in
many cases even when the basic assumptions no longer hold. The new methods are useful
mostly when photometric errors are small, the distribution of intrinsic colours is well
known, or one has \emph{prior} knowledge of the small-scale structures.
}


\maketitle

\section{Introduction}

Extinction measurements with near-infrared (NIR) observations are one of the main
methods of measuring the column density of dense interstellar clouds. Compared to optical
wavelengths, NIR optical depths are lower, enabling the observations to probe a wide
range of column densities. With the all-sky 2MASS survey \citep{Skrutskie2006}, the
structure of any nearby cloud can be studied in regions with $\sim 1-20$\,magnitudes of
visual extinction, reaching a resolution down to $\sim 1\arcmin$, depending on the
stellar density. The resolution and dynamical range of NIR observations make them
particularly relevant for studies of star formation from the formation of molecular
clouds to the structure of individual pre-stellar cores.

Optical extinction maps are  often based on star counts \citep[see,
e.g.,][]{Dobashi2005_DSS}, and the same method can be applied to NIR observations in
regions of higher optical depth. However, if data are available for several NIR bands,
better results are obtained by making use of the colour excesses of individual stars
\citep[e.g.,][and references below]{Cambresy1997_DENIS}. Because the spectral classes of
individual stars are usually not known, each extinction measurement requires an average
over a sufficient number of stars to overcome the statistical scatter of intrinsic
colours. The effective spatial resolution is thus dictated by the surface number density
of stars. However, for a sufficient sample of randomly selected stars, the uncertainty of
the average colour is small \citep[e.g.,][]{Lombardi2001_NICER, Cambresy2002_2MASS,
Davenport2014}. The average stellar colours vary over the sky, depending on the
contributions of different stellar populations. However, this can be taken into account
by using a nearby, extinction-free field as a reference or by using models that give
predictions of these variations \citep[e.g.,][]{Robin2014}. The reddening of stellar
radiation depends on the interstellar dust particles within the intervening clouds. The
NIR extinction curve is observed to be relatively constant between regions of different
density and Galactic location \citep[e.g.,][]{Cardelli1989, Wang2014}. Thus, the
difference between the observed and the intrinsic colours should result in reliable
estimates of the intervening dust mass. 

The extinction of several large areas has already been mapped using the 2MASS survey.
The studied clouds include the Polaris Flare \citep{Cambresy2001}, the Pipe nebula
\citep{Lombardi2006_Pipe}, Ophiuchus and Lupus clouds \citep{Lombardi2008_Oph}, Taurus
\citep{Padoan_2002_Taurus, Lombardi2010_Taurus}, Orion \citep{Lombardi2011_Orion,
Lombardi2014}, and Corona Australis \citep{Alves2014_CrA}. Most studies have used the
optimised multi-frequency method NICER \citep{Lombardi2001_NICER} that combines the
information of the $J-H$ and $H-K$ colours and uses Gaussian weighting to transform the
extinction estimates of individual stars into continuous extinction maps. The method
includes a so-called sigma-clipping method to filter out outliers like unextincted
foreground stars. All methods using averages of {\em observed} stars exhibit some bias
that is related to the column density variations on
scales below the size of the kernel
that is used to average measurements of individual stars. The extinction is
underestimated because fewer stars are observed through higher column densities. The
NICEST method \citep{Lombardi2009_NICEST} uses the scatter of extinction estimates to
make a statistical correction for this bias. Thus NICEST can result in higher and less
biased estimates at the locations of local column density maxima. This was also observed
in \citet{JuvelaMontillaud2015a}, which presented all-sky extinction maps based on 2MASS
survey and both the NICER and NICEST methods.

In addition to foreground stars, galaxies can represent a significant source of
contamination in the sample of background stars. The intrinsic colours of galaxies are
typically redder and, if this is not taken into account, will lead to higher extinction
estimates. Part of the galaxies can be removed based on their resolved size. Instead of
rejecting them altogether, they can be used for a separate extinction calculation. When
combined with the evidence of the stars, this can result in better extinction estimates,
especially at high Galactic latitudes where there are fewer stars
\citep{Foster2008_GNICER}.

In this paper we present a new method of calculating extinction maps based on NIR colour
excesses. 
The method employs a discretised version of the 2D intrinsic
colour distribution and uses Markov chain Monte Carlo (MCMC) methods
to estimate the probability distribution of extinction for each map
pixel.
We also
examine the possible benefits of considering the statistical or more direct
information on sub-beam column density variations, detection thresholds, and the
variation in intrinsic colour distributions as a function of apparent magnitude.

The structure of the paper is as follows. The methods implemented in a program
SCEX
(Star Colours to EXtinction) are described in Sect.~\ref{sect:methods}, and the synthetic
observations used in the tests are discussed in Sect.~\ref{sect:synthetic}.
Section~\ref{sect:results} presents the analysis of synthetic observations, using the
different variations of the basic method. We start with synthetic observations with
simple intrinsic colour distributions (Sects.~\ref{sect:T1}-\ref{sect:T2}) before using
stars simulated with the Besan\c{c}on model \ref{sect:1MMX}. In Sect.~\ref{sect:example}
we present an application to real data and compare the results to NICER and NICEST
extinction maps. We discuss the results in Sect.~\ref{sect:discussion} before listing
the final conclusions in Sect.~\ref{sect:conclusions}.

\section{Methods}  \label{sect:methods}

Extinction estimation with colour excess methods is based on the difference between the
observed and the intrinsic colours of a star. For example, the observed $J-H$ colour
(i.e., the magnitude difference $m_{\rm J}-m_{\rm H}$) becomes
\begin{equation}
   J-H  =   (J-H)_{\rm 0} + A_J - A_H,
\label{eq:ce}
\end{equation}
where $(J-H)_{\rm 0}$ is the intrinsic colour of the star, and $A_J$ and $A_H$ are the
amount of extinction in the two bands. If the intrinsic colour is known, Eq.~\ref{eq:ce}
gives the difference in the extinction between the two bands. The relation between
$A_J-A_H$ and the dust column density depends on dust properties. More directly, the
knowledge (or assumption) of the shape of the extinction curve enables conversions
between $A_J-A_H$ and the extinction in any single band. 

In this paper we discuss
observations in three NIR bands and two colours, $J-H$ and $H-K$. Thus, each star can be
represented as a point in a $(J-H, H-K)$ plane where the extinction by intervening dust
clouds moves the stars towards the upper right, in a direction determined by the shape of
the extinction curve and a distance determined by the amount of extinction. 
We do not consider here the uncertainties of the extinction curve. Because the spectral
classes of individual stars are typically not known, the extinction calculations need to
make assumptions about the probability distribution of the intrinsic colours. This is a
major source of uncertainty. To bring down the errors in the final extinction estimates,
one averages the information provided by a number of stars. For example, the NICER
method approximates the intrinsic colour distribution with a 2D normal distribution,
calculates least squares extinction values for individual stars, and calculates the
final estimate as an average of these values weighted by a Gaussian beam. 

The program SCEX uses a discretised presentation of the intrinsic colour distribution.
We need a probability distribution of the intrinsic colours $(J-H)_0$ and $(H-K)_0$.
This can be derived from observations of an extinction-free OFF field or from stellar
models. Unlike the NICER method, we do not describe the intrinsic colours using a
covariance matrix but directly discretise the $((J-H)_0,(H-K)_0)$ distribution onto a
2D array $P_C$. In this paper, this is based on a catalogue of simulated sources, the
number of objects per cell directly giving the relative probability of the corresponding
intrinsic colours. In practice, some smoothing of the $P_C$ distribution is required
because of the finite number of reference stars and because a sharp drop of $P_C$ to zero
probability would cause problems for MCMC. 
We use simple Gaussian smoothing with FWHM corresponding to 0.1 units in both colours.
This works adequately in all cases, since the $P_C$ distributions are generated with a large
number of sources, $\sim$40~000.

We quantify column density using $J$ band extinction, $A_J$. When $A_J$ is greater than
zero, the observed colours $J-H$ and $H-K$ move along the reddening vector defined by
$A_J-A_H$ and $A_H-A_K$, which in turn depends on the assumed dust properties. In
the simplest form of the method, we calculate extinction maps pixel by pixel, 
using $A_J$ as the only free variable. For an estimate of $A_J$, we calculate the 
dereddened colours of all sources close to the pixel. From the resulting location 
of each source in the colour-colour diagram, the discretised colour probability $P_C$ is 
used to estimate the probability value $P_i$ that the dereddened colours equal the 
intrinsic colours of the source. To get the probability of the $A_J$ value of 
a pixel, we calculate a weighted sum of the probabilities of individual sources $P_i$:
\begin{equation}
\ln P = \frac{ \sum W_i \ln P_i}{ \sum W_i}.
\label{eq1}
\end{equation} 
The weights $W_i$ include both spatial weighting and weighting with photometric errors:
\begin{equation}
W_i = W_{\rm{S,}i} \times W_{\rm{P,}i}.
\label{eq:weights}
\end{equation} 
The spatial weighting corresponds to spatial convolution, which depends on the
selected FWHM value of the extinction map and the distance between the pixel centre and 
the source, $r_i$:
\begin{equation}
W_{\rm{S,}i} \propto  \exp 
\left( \frac{ -4 \ln(2) r_i^2}{FWHM^2} \right).
\label{eq:WS}
\end{equation} 

Exact handling of photometric errors is possible but computationally expensive,
especially in the framework of the MCMC method adopted here. We employ a simplified version
where the photometric errors of individual sources are taken into account only
approximately. We follow Eq. 5 of \citet {Lombardi2001_NICER} where the variances of the
photometric errors $\sigma_j^i$ and the variance of the intrinsic colour distribution
$(\sigma_j)^2$ are used as follows:
\begin{equation}
W_{\rm{P,}i} \propto  
1 / \sum_j \left( (\sigma_j^i)^2 + (\sigma_j)^2 \right).
\label{eq2}  
\end{equation} 
This simplifies computations because only $A_J$, one value per pixel, is kept as a free 
variable. This weighting is not optimal because it does not retain information about the
relative uncertainties or covariances of the $J-H$ and $H-K$ colours. This is not a 
significant source of error in our test cases where the error estimates are smooth 
functions of magnitude, and the final uncertainty is dominated by the dispersion of 
intrinsic colours. However, in this procedure we also approximate the variance of 
intrinsic colour distribution with a single number even though the actual distribution
is not normal and varies depending on the location in the colour-colour plane. These 
simplifications speed up the calculations but can lead to sub-optimal results and will 
affect the width of the posterior probability distributions. However, we have verified 
that in simple cases the results remain essentially the same when using the full model 
where the errors of individual magnitude measurements are included as free variables.


The adopted simplified method can be seen as a mixture of NICER-like weighted mean
analysis and full likelihood analysis. In principle, we should marginalise the
probability over all possible values of intrinsic source colours, weighted by $P_C$, and
consider the individual uncertainties in each source's photometry. We are effectively
ignoring the photometric errors in the marginalisation over the intrinsic colours. We
then include the photometric uncertainties only in an approximate fashion, downweighting
points with large photometric uncertainties in Eq.~\ref{eq1}.

We have carried out the calculations using MCMC with the Metropolis algorithm.
During the MCMC calculation, $A_J$ is updated using random steps that are generated from
a normal distribution $N(0,\delta)$ and are accepted if the old probability $P(A_{J}
^{\rm Old})$ and the new probability $P(A_{J}^{\rm New})$ fulfil the criterion
\begin{equation}
\ln  P(A_{J}^{\rm New}) -  \ln P(A_{J}^{\rm Old}) > \ln(u),
\end{equation}
where $u$ is a uniform random number between 0.0 and 1.0. The step size $\delta$ is
adjusted to keep the acceptance rate at a level of a few tens of percent. 
We are maximising the posterior probability
\begin{equation}
P(A_{J}|{\rm data}) = \frac{P({\rm data}|A_{J}) \times P(A_{J})}{P({\rm data})},
\end{equation}
which makes it possible to include {\em \emph{a priori}} information of the $A_{J}$ values
or their distribution. However, in the simplest case of flat prior distributions
$P(A_{J}),$ this reduces to maximisation of the probability of Eq.~\ref{eq1}. 

Each pixel of the computed extinction map corresponds to an average estimate over one
Gaussian beam with the given FWHM. In the following examples, we adopt a pixel size of
1$\arcmin$ and a FWHM beam size of 3$\arcmin$. The quality of the extinction estimates
depends on the number surface density of stars, the accuracy of their photometry, and
the dispersion of their intrinsic colours. The MCMC calculations provide samples of
$A_J$ that together describe the probability distribution of this parameter. As the
final estimate shown in the maps, we use the median value of the $A_J$ samples.  
The number of samples used is a few thousand for the basic method and of the order of a
hundred thousand for the most complex ones.

We refer to the basic version of SCEX with a single free parameter per beam as {\bf
Method B}. We present below some modifications of this basic scheme that are then
tested in Sect.~\ref{sect:results}. We continue to refer to the sources as ``stars'',
although the data might more generally consist of both stars (possibly including
distinct populations) and galaxies. Calculations are carried out with the Markov chain Monte
Carlo (MCMC) method, but direct maximum likelihood estimation is a viable and
computationally faster alternative. Thus, the methods discussed in the following
sections are not limited to MCMC implementations. Similarly, although tests are
carried out using three NIR bands, the methods can be generalised to more channels,
although possibly with a steep increase in computational cost.

\subsection{Explicit {\em a priori} information on small scale structure} \label{sect:method_T}

Ideally one would have explicit information about the relative extinction at the location
of each star, i.e., of the cloud structure on scales below the beam size, so that one
could {\em specify} the difference $\Delta A_J^i=A_J^i - A_J$ between the extinction
to the star $A_J^i$ and the beam-averaged value $A_J$. This may sound like a drastic step, 
because one is trying to recover the extinction values $A_J$.
However, {\em \emph{a priori}} information often exists in some form. There may be dust
continuum observations that provide some information about mass distribution on small
scales. The extinction calculations themselves can reveal large-scale gradients that
imply that similar gradients are likely to also exist on smaller scales. A prescription
of relative extinction values within a beam does not affect the absolute values of the
$A_J$ estimates, and in particular, it will not directly influence the relative $A_J$
estimates calculated for different pixels. Thus, the resulting extinction map should, in
both scaling and morphology, be independent of the ancillary information about the small-scale structures. The basic method (B) is, of course, a special case where true extinction
is assumed to be constant over the beam.

We propose an improved version of SCEX, hereafter {\bf Method T}, where, in 
the calculation, we directly use the expected ratios $k_i$ between the extinction at the 
location of a star $i$ and the extinction averaged over the beam, $k_i=A_J^i/A_J$. The 
factors $k_i$ are specific to each star, but when a star is included in several beams, the 
values are independent between the beams. The method could be refined further by including 
an error distribution for the $k_i$ factors, but this was not investigated.
In the following, {\bf Method T1} refers to calculations that assume perfect knowledge of
sub-beam structure. (More precisely, exact value of $k_i$ is known for each observed
star.) {\bf Method T2} refers to a case where we do not use any ancillary information, and the
values of $k_i$ are based on a version of the extinction map itself.

\subsection{Compensating for bias} \label{sect:method_D}

Sub-beam structures are known to bias estimates of beam-averaged extinction because
one is more likely to detect stars through the low column density parts of the beam. The
initial results of Method~B also show the effects of sub-beam structures in the
$(J-H,H-K)$ plane, where the individual dereddened colours do not coincide with the
maximum probability along the reddening vector but are scattered around the maximum over
an area much larger than expected based on photometric errors alone. The recovered $A_J$
value should be an estimate of the beam-averaged extinction, but it is not a very good
estimate of the extinction toward an individual star. The large scatter suggests that
the sub-beam variations should be considered as a part of the model.

We define here{\bf Method D1}, a new variant of SCEX, which aims to compensate for this
bias. The strategy is to include the probability distribution of the extinction of an
individual {\em detected} star, $p(A_J^i)$, in the model. We derive $p(A_J^i)$ from two
elements: the relative fluctuations of extinction $A_J^i/A_J$ and the expected
distribution of stars as a function of magnitude. In the case of no extinction, we model
the cumulative number of stars brighter than $m_J$ as $n \propto 10^{\alpha \times
m_J}$\citep[cf.][]{Cambresy2002_2MASS}. The parameter $\alpha$ can be obtained from
observations of an extinction-free OFF region. We assume that the relative
fluctuations $A_J^i/A_J$ follow a log-normal distribution with $\sigma \sim 0.5$. This
leads to 
\begin{equation}
p(\Delta A_J^i) \propto   N(\ln \/ (A_J^i/A_J), \sigma) \times 10^{-\alpha \Delta A_J^i}.
\label{eqX}
\end{equation}
Because the detection probability is lower for stars with higher values of $A_J^i$, the
distribution is skew, reflecting that the true beam-averaged extinction 
is likely to be higher than the average extinction towards the detected stars.

In the calculation, the free variables are the beam-averaged estimates $A_J$ and the
differences $\Delta A_J^i=A_J^i-A_J$ for each star $i$ within the beam. The probability
consists of the two terms mentioned above, $P_C$ and $p(\Delta A_J^i)$:
\begin{equation}
p \propto 
\prod_i 
P_{\rm C}(m_j^i, A_J + \Delta A_J^i) \,
\exp \left[ -\frac{(\ln [(A_J+\Delta A_J^i)/A_J])^2}{2\sigma^2}    \right]
\, 10^{-\alpha \Delta A_J^i}.
\label{eq:combined}
\end{equation}
Here $P_{\rm C}$ corresponds to the probability of the extinction-corrected colours and,
therefore, only depends on the observed magnitudes $m_j^i$ ($j$ referring to the
presence of multiple bands) and the current extinction estimate $A_J+\Delta A_J^i$
of a star. The rest of the probability depends on the difference between the current
extinction estimates of individual stars, $A_J^i$, the current estimate of the
beam-averaged extinction, $A_J$, and the amount of sub-beam variations assumed,
$\sigma$. The above equation is written only as a proportionality because the correct
normalisation of Eq.~\ref{eqX} is not yet defined. In the program this is taken into
account explicitly by calculating values relative to the integral over the full
probability distribution (for the current value of $A_J$, see below).

To speed up the calculations, the probability distribution of $\Delta A_J^i$ is not
updated during the calculation even though it depends on the estimate of the
beam-averaged extinction, $A_J$. In practice, we start the calculation with NICER
estimates of $A_J$. SCEX is then run to estimate a new extinction map. If the extinction
values change, the probabilities $p(\Delta A_J^i)$ also change. To get a consistent
solution, the whole SCEX run must be repeated using the updated $p(\Delta A_J^i)$
values. After a couple of iterations the extinction estimates no longer change.

Because Method D1 increases extinction estimates based on the magnitude of the
$10^{-\alpha \Delta A_J^i}$ term, it shows some similarity to the NICEST
routine \citep[see Eq. 34 in][]{Lombardi2009_NICEST}. However, there are fundamental
differences. First, if the beam contains only one star, NICEST directly returns the
estimate calculated for this star (apart from a small correction dependent on the
magnitude uncertainty), while Method D1 returns a higher value: the extinction seen by a
random detected star is expected to be lower than the mean extinction. Second, Method
D1 requires assuming the extinction fluctuations to be able to
estimate the bias correction. This can be a major drawback, especially if the correction
is sensitive to the assumptions.


As an alternative, we implement as SCEX {\bf Method D2} as a correction that is more similar to
the NICEST algorithm.  We must again explicitly calculate independent extinction
estimates $A_J^i$ for each individual star. The values $A_J^i$ are averaged and 
weighted
them by the beam, the photometric uncertainties, and factors proportional to $10^{+\alpha
A_J^i}$ \citep[cf. Eq. 34 in][omitting the small second term]{Lombardi2009_NICEST}.
Thus, in Method D2 the values $A_J^i$ act as the free variables, and the beam-averaged
values $A_J$ are calculated on each step based on the current values of $A_J^i$. The
final extinction estimate is the median over the $A_J$ values on different MCMC steps.

The probability distribution of $A_J^i$ may have multiple peaks corresponding to
different source populations along the reddening vector. When stars are treated
independently, $A_J$ will also end up having multiple local maxima. If the solution
were constrained with the knowledge that all sources within the beam are affected by a
similar extinction, the result could be limited to the one solution that is consistent
with the evidence of all sources. Thus, {\bf Method D2} could be developed further by
including a term that depends in the probability calculation on the differences
between the individual and the beam-averaged extinction estimates, but this was not
investigated.

\subsection{Calculations with several sub-populations} \label{sect:method_P}

\begin{figure}
\includegraphics[width=8.8cm]{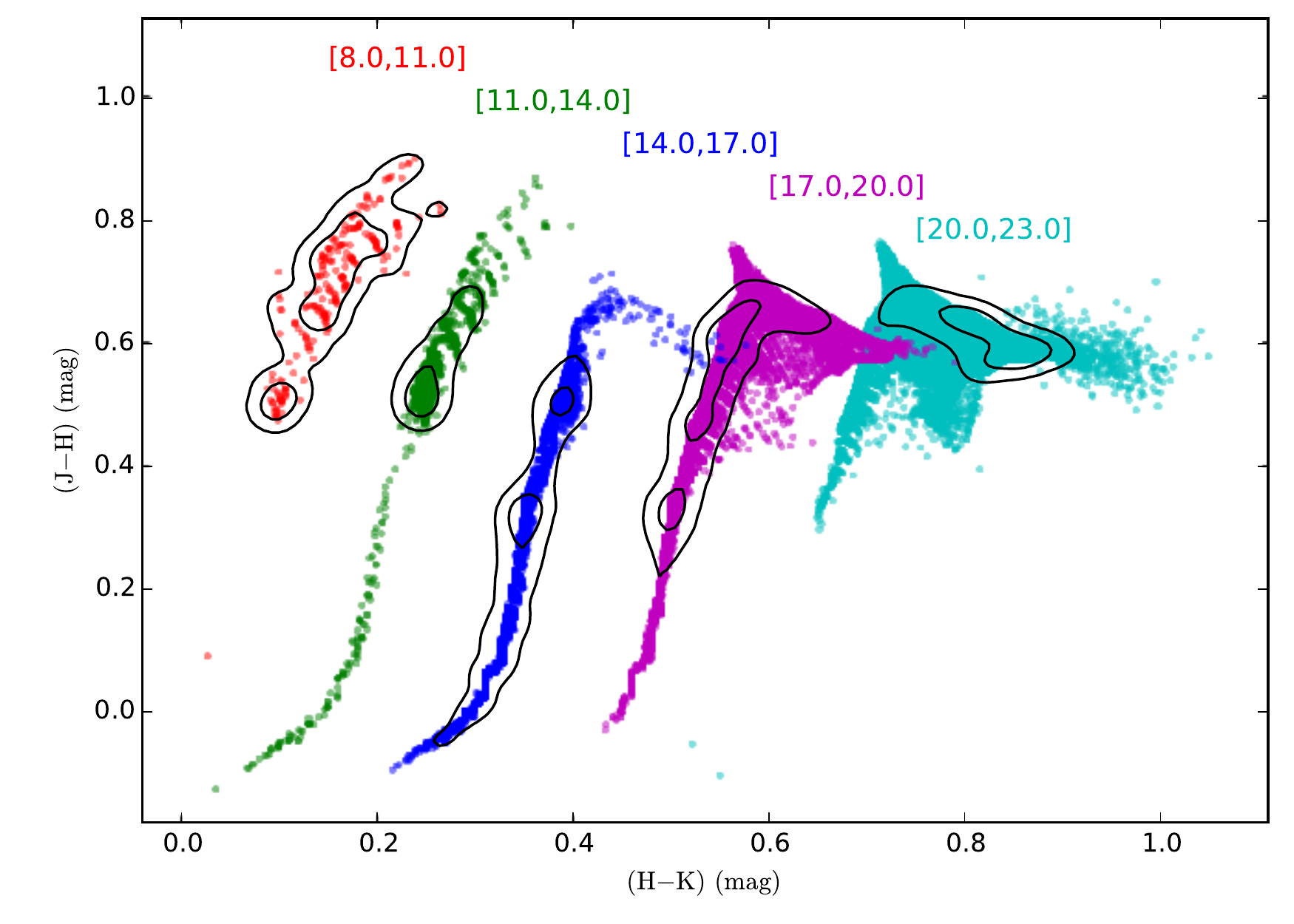} 
\caption{
Stellar colours for Galactic direction ($l$, $b$)=(10$\degr$, 20$\degr$) for a sample of
stars simulated from the Besan\c{c}on model \citep{Robin2014}. The
different intervals of $J$ band magnitude are plotted in different colours, as indicated
by the values in the upper part of the figure. After the first one (8.0-11.0\,mag), the
subsequent sets of points are shifted by $\Delta(H-K)=0.15$\,mag, for better
readability. The smoothed contours are drawn at 10\% and 50\% of the maximum star
density.
}
\label{fig:plot_BES}
\end{figure}

The final step is to consider the presence of several sub-populations of sources, each
with a different distribution of intrinsic colours. One example could be the separation
of stars and galaxies, if the classification can be done beforehand (for example, based
on source morphology). The idea can be expanded to different stellar populations that
could be defined beforehand (based on additional photometric or spectroscopic data) or
based on the apparent magnitudes. In the latter case, the observed stars would be
corrected for the estimated extinction and their colours compared to the probability
distribution for stars of similar apparent brightness. Figure~\ref{fig:plot_BES} shows
the expected colours for different magnitude intervals in the direction ($l$,
$b$)=(10$\degr$, 20$\degr$), as calculated from the Besan\c{c}on model. This illustrates
the possible correlations that exist between the apparent magnitudes and the intrinsic
colours.

If the sources are pre-classified, the calculation is straightforward. Instead of a
single distribution of intrinsic colours, one uses for each source the probability
distribution appropriate for that source category. If the classification is based on
apparent magnitudes corrected for extinction, the classification depends on the $A_J$ 
values that are being calculated. Therefore, either the shifts between
`populations' (magnitude bins) need to become part of the MCMC process itself or the
whole calculation needs to be iterated. In the implemented SCEX {\bf Method P}, we
opt for the second alternative because this is in practice faster. When started with
NICER estimates, the $A_J$ values do not change very significantly, and calculations are
practically converged after the second iteration.

The different variations of SCEX are summarised in Table~\ref{table:methods}.

\begin{table}
\caption{Alternative methods implemented in the SCEX program}
\label{table:methods}
\begin{tabular}{ll}
\hline
Method   &  Description  \\
\hline
B        &  The basic program \\
T1       &  Using perfect knowledge of relative extinction on \\
         &  small scales, below the beam size \\
T2       &  Using the calculated extinction values to  \\
         &  estimate relative extinction on small scales \\
D1       &  Using a probability distribution $P_C$ modified \\
         &  to compensate for effects of a detection threshold \\
D2       &  Using weighting $\propto 10^{+\alpha A_J^i}$ to compensate \\
         &  for effects of a detection threshold \\
P        &  Method using {\em \emph{a priori}} classification of sources \\
\hline
\end{tabular}
\end{table}

\section{Synthetic observations} \label{sect:synthetic}

We used simulated test data that describe the intrinsic stellar colours, the photometric
errors of measured magnitudes, the spatial distribution of extinction, and the spatial
distribution of stars. The input data are in the form of synthetic catalogues of $J$, 
$H$, and $K$ magnitudes with error estimates and images of the true extinction.
We have extracted a catalogue of 2MASS stars from a sky region free of significant
extinction. This catalogue is only used to determine completeness curves and relations
between observed magnitudes and photometric errors. The distribution of intrinsic
colours is set either using {\em \emph{ad hoc}} distributions (Sects.\ref{sect:T1} and
\ref{sect:T2}) or the Besan\c{c}on stellar population model of \citet {Robin2014} (Sect.
\ref{sect:1MMX}). For example, in the first tests (Sect. \ref{sect:T1}), we used simple 2D
Gaussian distributions in the ($H$-$K$, $J$-$H$) plane. A simulated catalogue of stars
was created randomly according to the adopted distribution of intrinsic colours
and used to define the statistics of unextincted stars (the reference field), the
scatter of the intrinsic stellar colours, and the correlations between $J$-$H$ and
$H$-$K$ values. In the case of NICER this results in estimates of the average colours and
their covariance matrix. In the case of SCEX it results in a discretised 2D map of
probability $P_C$ in the colour-colour space (see Sect.~\ref{sect:methods}).

The spatial distribution of extinction in the ON field is described by an $A_J$ map with
1$\arcmin$ pixels. In most cases, we used a map derived from Herschel dust emission
measurements \citep[field G4.18+35.79 in][]{GCC-V} but scaled to a different value of
maximum $A_J$. The data cover a $\sim 33\arcmin \times 33\arcmin$ map with 1$\arcmin$
pixels. We used an extinction curve with relative extinctions $A_H/A_J$=0.64 and
$A_K/A_J$=0.40, which corresponds to Galactic dust with $R_{\rm V}=3.1$
\citep{Cardelli1989}. In the ON field, we generated stars at random locations, applied
extinction according to the input $A_J$ map (assuming that all stars are located behind
the cloud), and added normal-distributed noise to magnitude measurements according to the
magnitude-dependent error curves. Finally, we removed stars stochastically, according to
the completeness curves. The photometric errors may be scaled with a constant $k_{\rm
noise}$, the value $k_{\rm noise}$=1 corresponding to typical errors in the 2MASS
survey. We kept only those stars that are detected in all three bands. The completeness in
$J$ and $H$ bands drops in the default case around $\sim$14\,mag and in $K$ band around
$\sim$13\,mag, according to the completeness curves derived from 2MASS data.

\section{Results} \label{sect:results}

In this section we present results from tests carried out with a synthetic column
density map, {\em \emph{ad hoc}} distributions of intrinsic colours, and distributions of
magnitudes and magnitude errors extracted from 2MASS data of a reference field (see
Sect.~\ref{sect:synthetic}). The peak extinction is scaled to a value of $A_J=$2.5\,mag
at 1.0$\arcmin$ resolution. At the 3.0$\arcmin$ resolution of the calculated extinction
maps, the peak value is $A_J=1.8$\,mag ($A_V=6.4$\,mag for $R_V=3.1$). Unless
otherwise stated, the number of detected stars in the map was set to 5000.

\subsection{Gaussian-distributed reference colours} \label{sect:T1} 

\begin{figure}
\includegraphics[width=8.8cm]{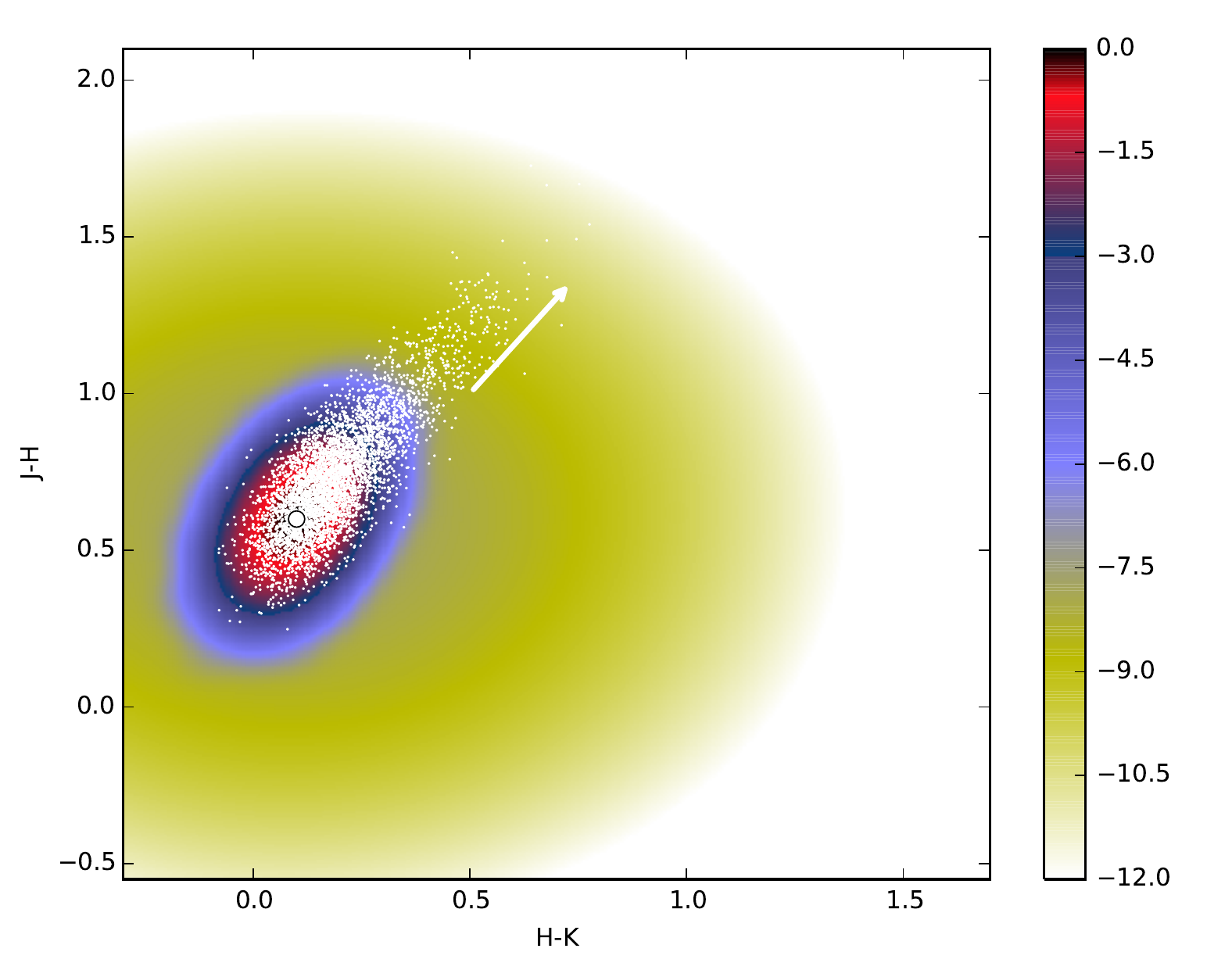} 
\caption{
Probability distribution of reference colours (colour image and logarithmic colour bar)
in a test with a single Gaussian distribution of intrinsic colours. The white arrow
points in the direction of the reddening vector, the length corresponding to $A_J=1.0$\,mag
of extinction. The white dots show one realisation of the reddened stars in the ON
field.
}
\label{fig:T1_00_info}
\end{figure}

We started with a test where the reference colours are distributed according to a
two-dimensional Gaussian (see Fig.~\ref{fig:T1_00_info}), which perfectly fits the
assumption that reference colours can be characterised using a covariance matrix. In
SCEX we in principle also have perfect knowledge of the intrinsic colours. However, in
practice the probability distribution $P_C$ is estimated based on a simulated catalogue of
some 30~000 unreddened stars. The discretised probabilities $P_C$ were calculated as
described in Sect.~\ref{sect:methods}, the discretisation introducing some noise and the
convolution of $P_C$ slightly smoothing the probability distribution. Thus, in this
particular case, the covariance matrix should provide a more accurate description of the
probabilities.

Figure~\ref{fig:T1_00_info} shows the probability distribution used as input for the
MCMC method, together with one realisation of the reddened star colours in the ON field
($k_{\rm noise}$=0.3). The number of stars above detection threshold was 5000, and
taking the map size of $\sim 33\arcmin$ into account, the average stellar density is around five stars per pixel (on average $\sim 47$ star per solid angle of the
$FWHM=3.0\arcmin$ Gaussian).  

\begin{figure}
\includegraphics[width=8.8cm]{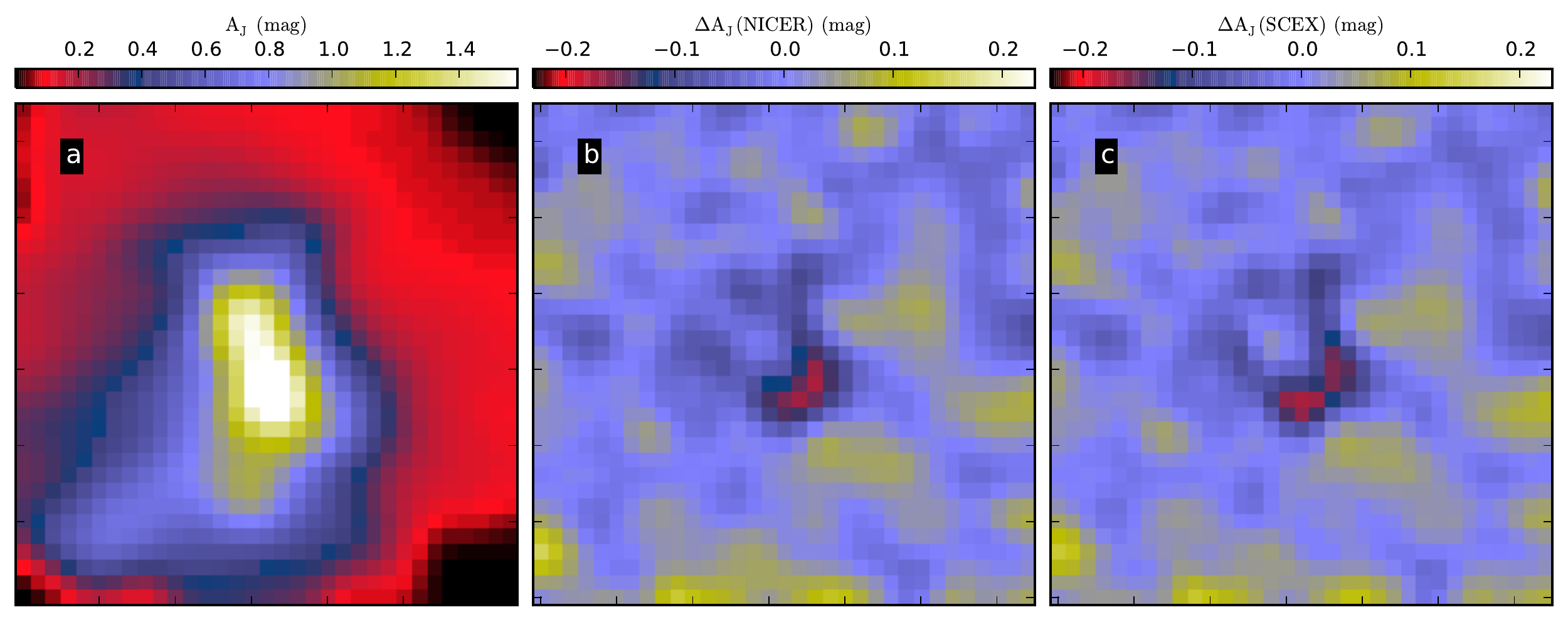} 
\caption{
Results of a test with a single Gaussian distribution of intrinsic colours. The leftmost
frame shows the input map of $A_J$ that is convolved to 3$\arcmin$, the resolution of
the calculated extinction maps. The other frames show the difference between the
estimated and the true extinction values for NICER (frame b) and SCEX Method
B (frame c). The negative values around the column density peak indicate bias that is
related to extinction gradients.
}
\label{fig:T1_00_3dmap}
\end{figure}

\begin{figure}
\includegraphics[width=8.8cm]{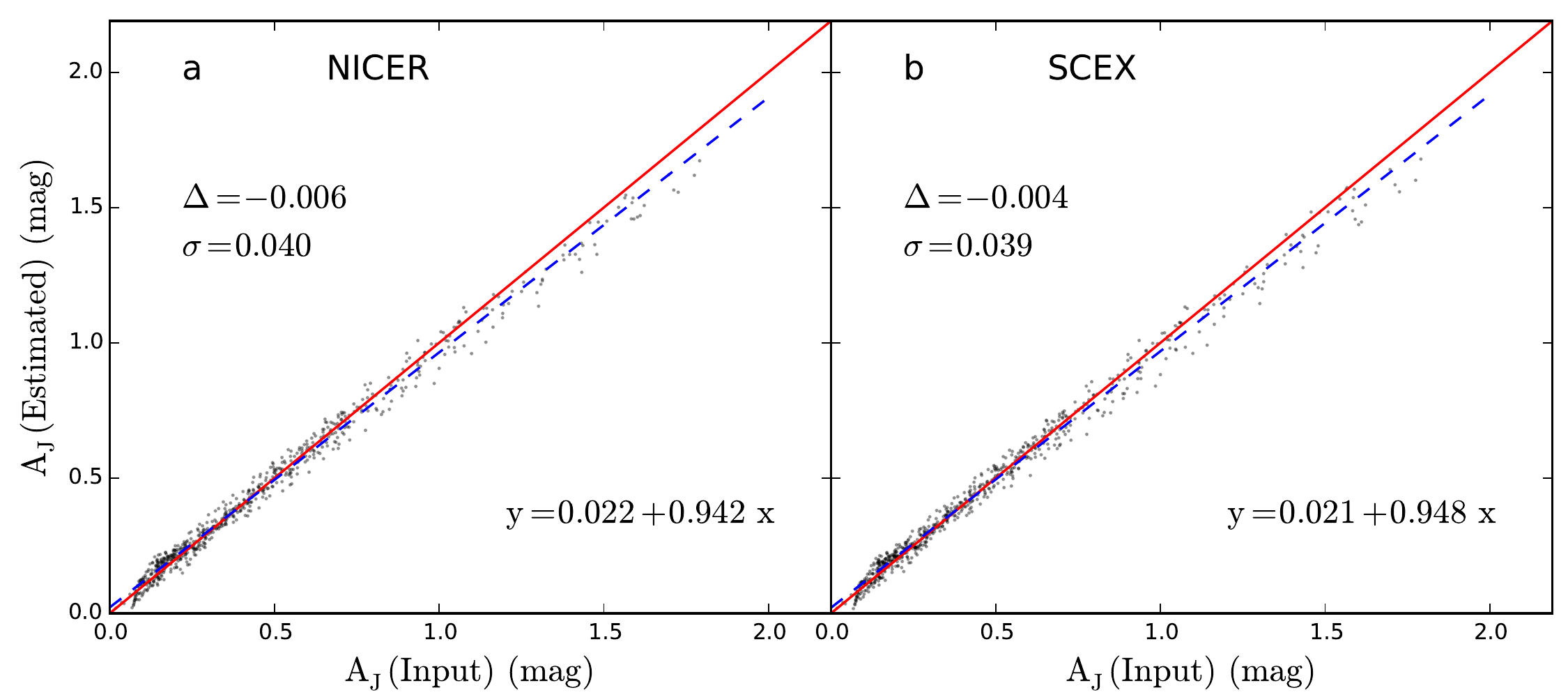} 
\caption{
Results of Fig.~\ref{fig:T1_00_3dmap} as scatter plots and as NICER and SCEX estimates as
functions of the input $A_J$ (convolved to 3$\arcmin$). The frames include values of the
bias $\Delta$ ($\Delta=\langle A_J^{\rm Estimate}\rangle-\langle A_J^{\rm
Input}\rangle$) and the rms error $\sigma$.
}
\label{fig:T1_00_scat}
\end{figure}

Figures~\ref{fig:T1_00_3dmap} and \ref{fig:T1_00_scat} show the input map and the errors
in the extinction maps calculated with NICER and Method B.
In the figures we also quote
the bias, the mean difference between the estimated and the true extinction values, 
$\Delta=\langle A_J^{\rm Estimate}\rangle-\langle A_J^{\rm Input} \rangle$. The residual
maps in Fig.~\ref{fig:T1_00_3dmap} are very similar between NICER and SCEX. The maximum
error is slightly larger for SCEX, but the rms error and bias are very similar (see
Fig.~\ref{fig:T1_00_scat}). The results vary only a little from one realisation to the
next. The errors are always of similar magnitude, but the order of the two methods can
also change, possibly because of the remaining stochastic noise in SCEX estimates.

\subsection{Complex distribution of reference colours} \label{sect:T2} 

\begin{figure}
\includegraphics[width=8.8cm]{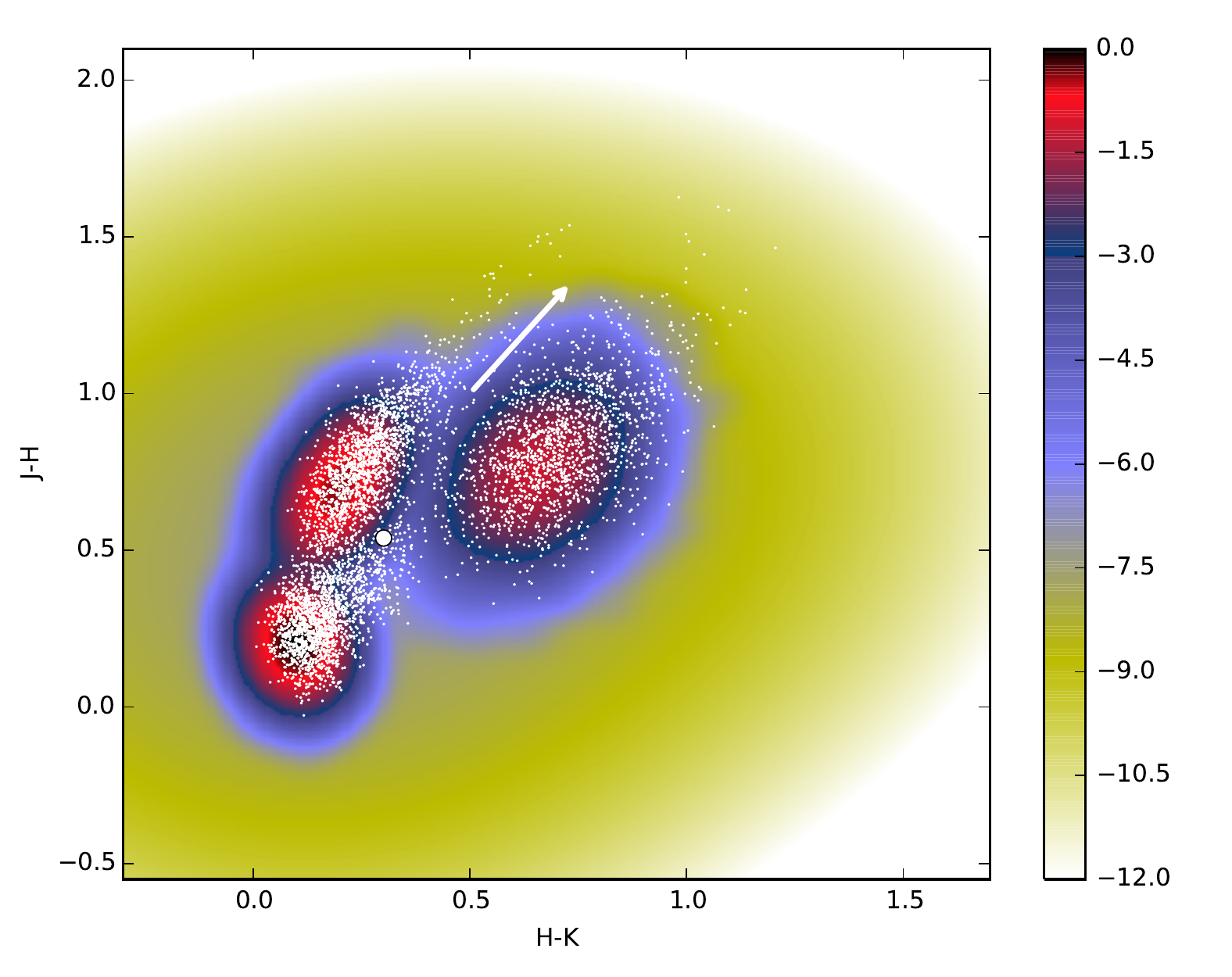} 
\caption{
Probability distribution of reference colours (colour image and logarithmic colour bar)
in the second test case. The white arrow points in the direction of the reddening vector,
the length corresponding to an extinction of $A_J=1.0$\,mag. The white dots show one
realisation of reddened observed colours. The larger white circle indicates the mean
intrinsic colours.
}
\label{fig:1B_info}
\end{figure}

SCEX should present some advantages if the distribution of intrinsic colours is more
complex than the single 2D Gaussian of the previous test (Sect.~\ref{sect:T1}). To
examine this in practice, we simulated intrinsic colours consisting of three Gaussian
distributions. The distribution is basically {\em \emph{ad hoc}} but has some resemblance to a
mixture of stellar populations and galaxies with redder colours. The example is chosen
explicitly to represent a case where the differences between the methods should be
clear. The distribution of intrinsic colours and one realisation of stars in the ON
field ($k_{\rm noise}$=0.3) are shown in Fig.~\ref{fig:1B_info}. 

\subsubsection{The basic method} \label{sect:basic}

\begin{figure}
\includegraphics[width=8.8cm]{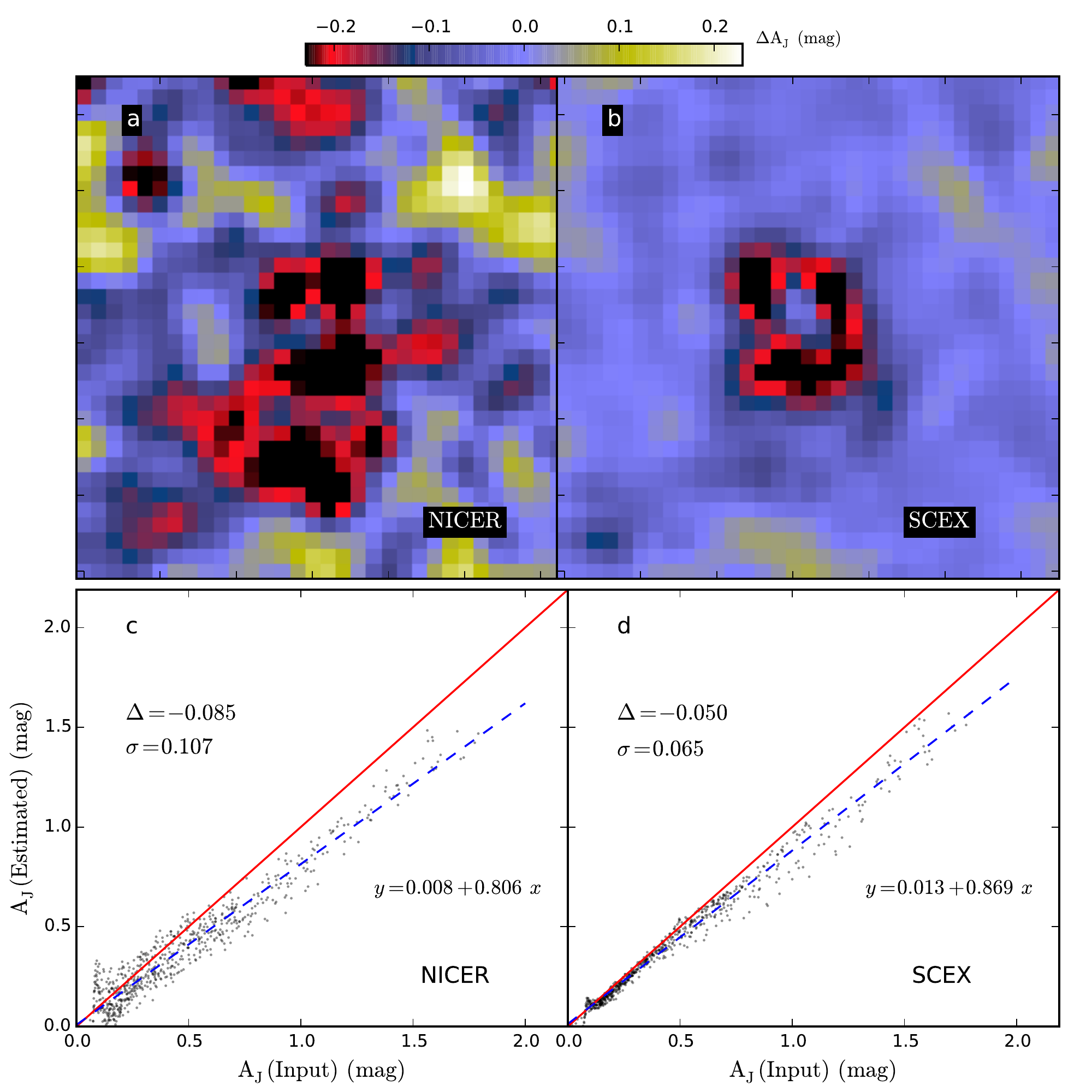} 
\caption{
Errors of NICER and SCEX (Method B) extinction estimates in test involving a complex
distribution of intrinsic colours ($k_{noise}=0.3$). The upper frames show the
difference between the calculated extinction maps and the input map convolved to the
same $3.0\arcmin$ resolution. Lower frames show the estimates as a function of the true
$A_J$ values. The numerical values of average bias $\Delta$ and rms errors $\sigma$ are
given in the figure.
}
\label{fig:1B_map_sca}
\end{figure}

Figure~\ref{fig:1B_map_sca} shows the NICER and Method B estimates in relation to the
true $A_J$ values of the simulation. Method B has smaller errors especially in low
column density regions, and the improvement can be attributed to the use of a more
accurate description of the intrinsic colour distribution. Both methods underestimate
the extinction around the column density peak. Calculated over the whole map, the rms
error and the bias of the SCEX map are some 40\% lower. Because the rms values are
strongly affected by the errors at high $A_J$, the differences are clearest below
$A_J\sim$1\,mag.

When noise is increased from $k_{\rm noise}=0.3$ to $k_{\rm noise}=1.0$, the difference
between the methods is reduced. This can be expected because added noise makes the
observed colour distributions more Gaussian. Considering the complexity of the simulated
intrinsic colour distribution, NICER performs quite well, although the overall rms
error of SCEX maps is still 20-25\% lower. The bias of Method B is also smaller but this
is a rather small effect, considering that in the colour-colour plane, the source
populations are separated by $A_J\sim 1$\,mag along the reddening vector. If the
noise is increased further, the distribution of intrinsic colours (including magnitude
errors) approaches a single Gaussian, and the difference between NICER and SCEX slowly
vanishes.

\subsubsection{Template of $A_J$ variations}  \label{sect:1T} 

The method T makes use of an input template of the $A_J$ variations on scales below the
beam size (see Sect.~\ref{sect:method_T}). In the first test we used the true $A_J$ map
at $1.0\arcmin$ resolution. The method needs the ratios of $A_J$ values along individual
lines of sight relative to the value in a map convolved to the resolution of the final
extinction map. In simulations the correct ratios can be easily extracted from the input
data, which represent perfect knowledge of all small-scale structure. This idealised
case indicates the maximum potential benefit from this method. As emphasised in
Sect.~\ref{sect:method_T}, the added information only concerns the relative extinction
values within each beam separately and does not directly affect either the absolute or
relative pixel values of the resulting $A_J$ maps.

\begin{figure}
\includegraphics[width=8.8cm]{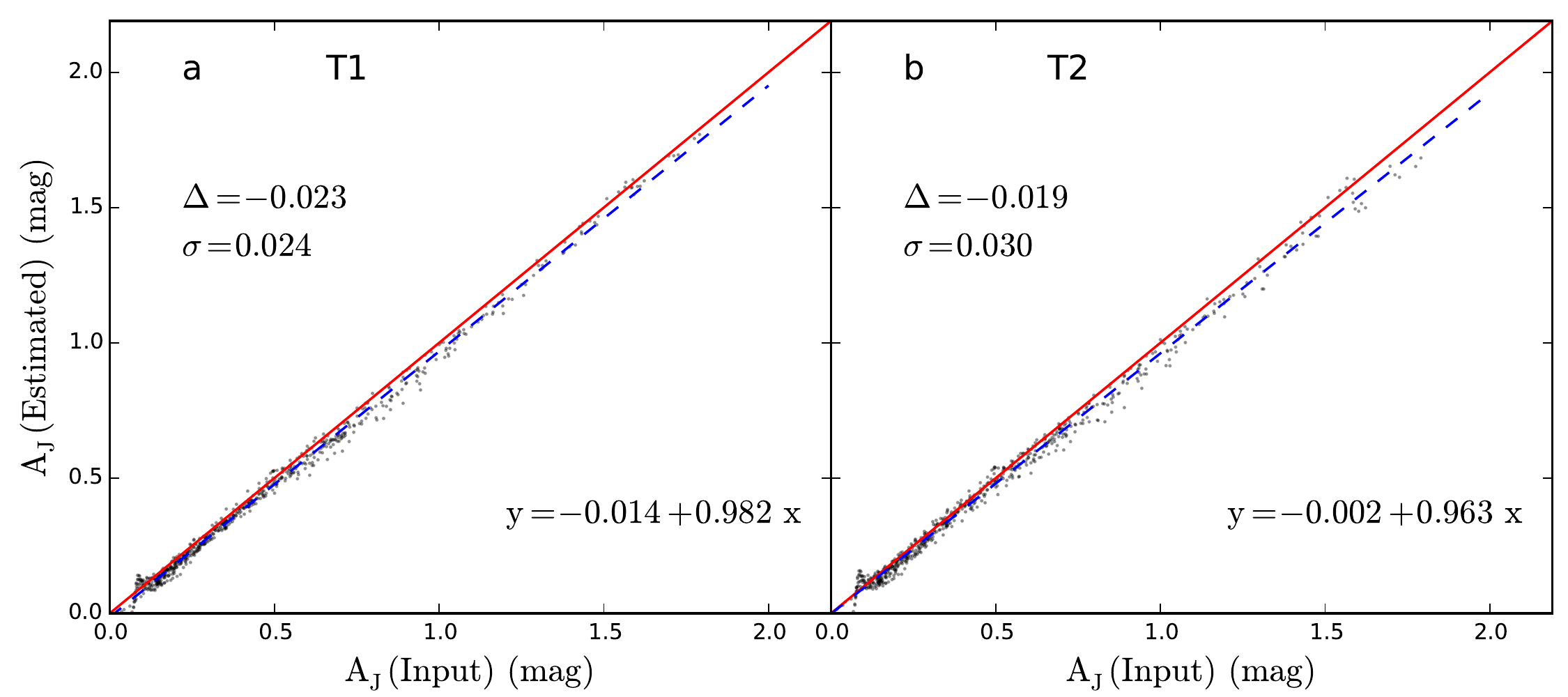} 
\caption{
SCEX Method T results plotted against the true values. SCEX calculations use
either accurate knowledge of small-scale $A_J$ variations (Method T1, frame a) or use
the NICER extinction map as a template of those variations (Method T2, frame b). The
average bias $\Delta$ and rms errors $\sigma$, as well as the parameters of least square
lines (estimates vs. true values) are given in the frames.
}
\label{fig:T1_T2}
\end{figure}

Figure~\ref{fig:T1_T2}a shows the that knowledge of the small-scale structure does
have a very significant effect on the accuracy of the extinction estimates. While the
rms error of NICER was $\sigma=0.11$, the errors of Method T are smaller by almost a
factor of $\sim 5$. This is also almost a factor of three below Method B errors in
Fig.~\ref{fig:1B_map_sca}. The systematic errors are almost completely removed, because the
least squares slope (between estimated and true values) is $\sim$0.99 compared to
0.87 for Method B and 0.81 for NICER.

We argue in Sect.~\ref{sect:method_T} that accurate information of sub-beam structure
could be available from observations at other wavelengths. However, we also consider the
possibility of using the $A_J$ map itself as the input template. In this case, the template
contains no information of real small-scale structure, but it does contain some
information of large-scale extinction gradients. In practice, we estimated extinction
variations with the NICER map, using the ratios between the values of individual pixels
and smoothed estimates obtained by convolving the same map with a beam of $3\arcmin$. In
other words, we are estimating the extinction variations using the ratio of $4.2\arcmin$
and $3.0\arcmin$ maps instead of the ideal case of $3\arcmin$ resolution vs. infinite
resolution. 

The result of the test with the NICER input map is shown in Fig.~\ref{fig:T1_T2}b. The
results are worse than when perfect knowledge of the small-scale structure was
available. However, the errors are only half of what they were for Method B (e.g., the
slope of the least squares fit is 0.96 for Method T and 0.87 for Method B). The errors
are greatest at the highest $A_J$ values, where Method T underestimates the true
extinction. This is natural because the template does not have the resolution to probe
$A_J$ variations at the peak, and therefore, the errors seen in
Fig.~\ref{fig:1B_map_sca} are only partially eliminated.

One might expect some improvement if using as input the corresponding deconvolved
extinction map that could predict larger sub-beam column density variations at the
location of the main peak. We tested this using a deconvolved NICER map (three
iterations with van Cittert algorithm). However, this resulted in no improvements. 
Although the deconvolved map appeared to give a good description of the most compact
structures, the bias of the highest $A_J$ values was not reduced. We went even further to
feed the deconvolved result of the extinction calculation in as the template map of
the next iteration but without any significant improvement.  It is still possible that a
deconvolved template map might work better in other circumstances, such as when higher
stellar density allows more reliable deconvolution.

\subsubsection{Corrections for bias} \label{sect:debias}

Apart from Method T, none of the above calculations make any correction for the fact
that the probability distribution of the extinctions of detected stars is not symmetric
with respect to the true beam-averaged extinction. In Method D1, we pre-calculated these
probability distributions based on the OFF field statistics and included them as part of
the calculation. The simulation is still based on the intrinsic colours shown in
Fig.~\ref{fig:1B_info}.

\begin{figure}  
\includegraphics[width=8.8cm]{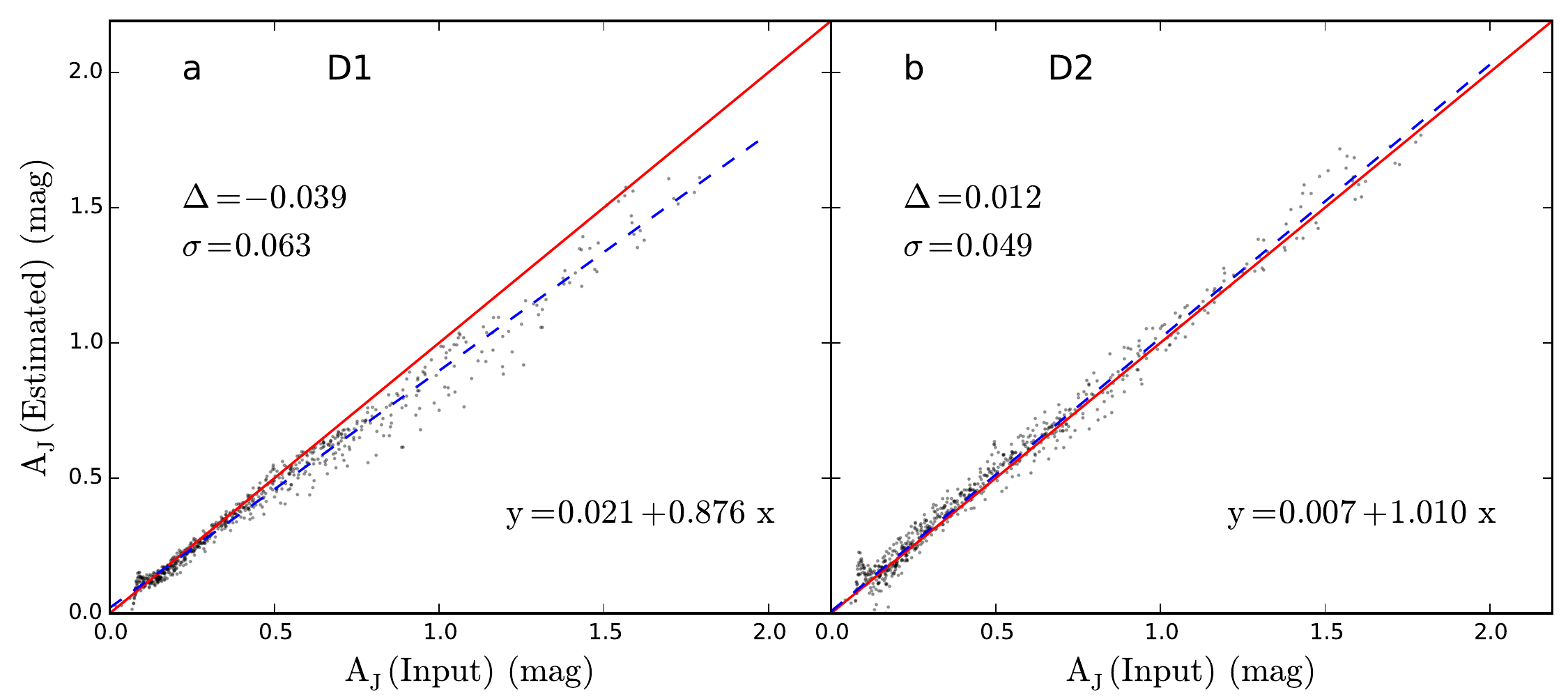} 
\caption{
Results of Methods D1 and D2 for the same case as in Fig.~\ref{fig:1B_map_sca}.
Calculations with Method D1 assume a log-normal model
with a standard deviation of $\sigma=0.5$ for column density fluctuations. 
}
\label{fig:D1_D2_scat}
\end{figure}

Figure \ref{fig:D1_D2_scat}a shows the result for Method D1 when the column density
fluctuations are assumed to follow a log-normal distribution with standard deviation
$\sigma=0.5$. Compared to Method B, the rms errors are slightly smaller, but
disappointingly the least squares slope between the estimated and true extinction values
has increased only marginally from 0.87 to 0.90. If the value of $\sigma$ is increased,
the slope gets systematically steeper, but the estimates become unbiased only with a much
higher value of $\sigma \sim $1.2.
Although the magnitude of the expected sub-beam fluctuations is not completely an {\em
\emph{ad hoc}} parameter, it cannot be determined directly from observations without higher
resolution data. Furthermore, the statistics of the fluctuations can change across a map
(e.g., between turbulence-dominated and gravitation-dominated regions). The column
density distribution of the map used in this test is in fact very far from the assumed
log-normal shape. On the high column density side, the distribution has a long power-law
tail that also extends down to values well below the mean column density. This partly
explains the poor performance of Method D1 in Fig.~\ref{fig:D1_D2_scat}a.

\begin{figure} 
\includegraphics[width=8.8cm]{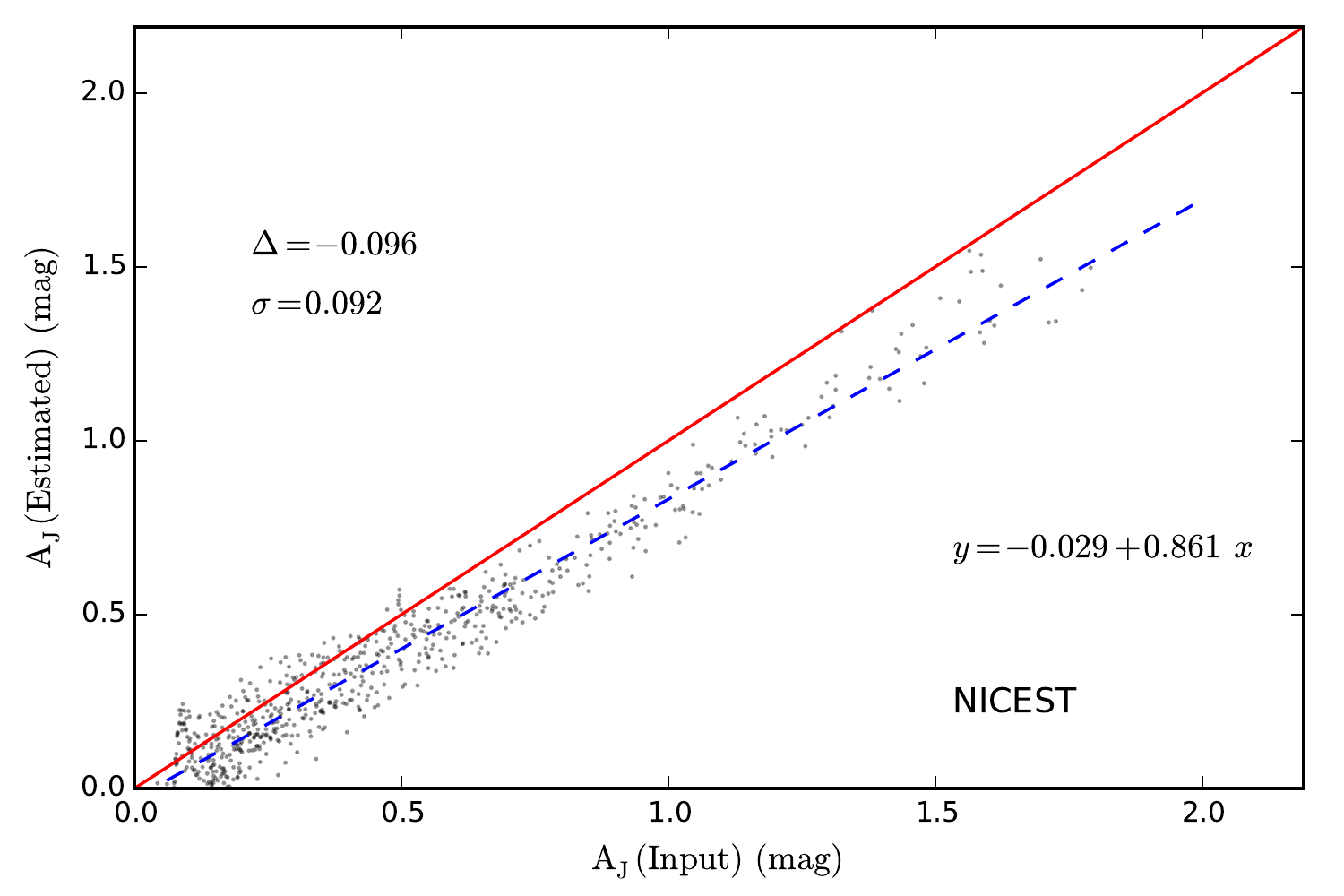} 
\caption{
NICEST result for the test same case as shown in Fig.~\ref{fig:D1_D2_scat}.
}
\label{fig:NICEST}
\end{figure}

Method D2 is essentially the same as the NICEST method, except for the use of
a discretised $P_C$ distribution.
For $\alpha$ we use a value $\alpha=0.31$ \citep[cf][]{JuvelaMontillaud2015a}.
In the case of Fig.~\ref{fig:D1_D2_scat}b, Method D2 shows practically no bias, and the
least squares slope between D2 estimates and true extinction values is 0.99. The rms
errors are lower than for Method B but higher than for Method T. In this case 
$A_J^i$ values of individual stars were calculated completely independently, making
the calculation quite straightforward. If there was more significant ambiguity in the
$A_J^i$ values (caused by multiple maxima in the intrinsic colour
distribution), it might still be beneficial to include penalty for the dispersion of
$A_J^i$ values within a beam. This would force all $A_J^i$ estimates to converge to a
consistent solution before the final beam-averaged extinction estimate is calculated.
Directly using NICEST with an assumption of Gaussian-distributed intrinsic colours
results in some improvement over NICER (see Fig.~\ref{fig:NICEST}) but still large bias
compared to Method D2.

\subsubsection{Case of two source populations} \label{sect:1MM}

As the first test of cases of multiple pre-classified source populations, we classified
the sources of Fig.~\ref{fig:1B_info} to two main components, with the sources found around
($H-K$, $J-H$)=(0.8,1.0) forming one component. The classification is assumed to take place
before SCEX calculation, based on external information. We also assume that the
intrinsic colour distributions are known for both populations separately. In practice,
we use the same input catalogues as before but tag each source based on its known
category. Because the generated source populations are well separated in colour-colour
space, the noise of the two $P_C$ arrays is similar to the noise in the previous single
$P_C$ array. The calculations of Method P correspond to Method B except for the use of
two $P_C$ arrays. In particular, no steps are taken to correct for the bias (cf.
Sect.~\ref{sect:debias}).

\begin{figure}
\includegraphics[width=8.8cm]{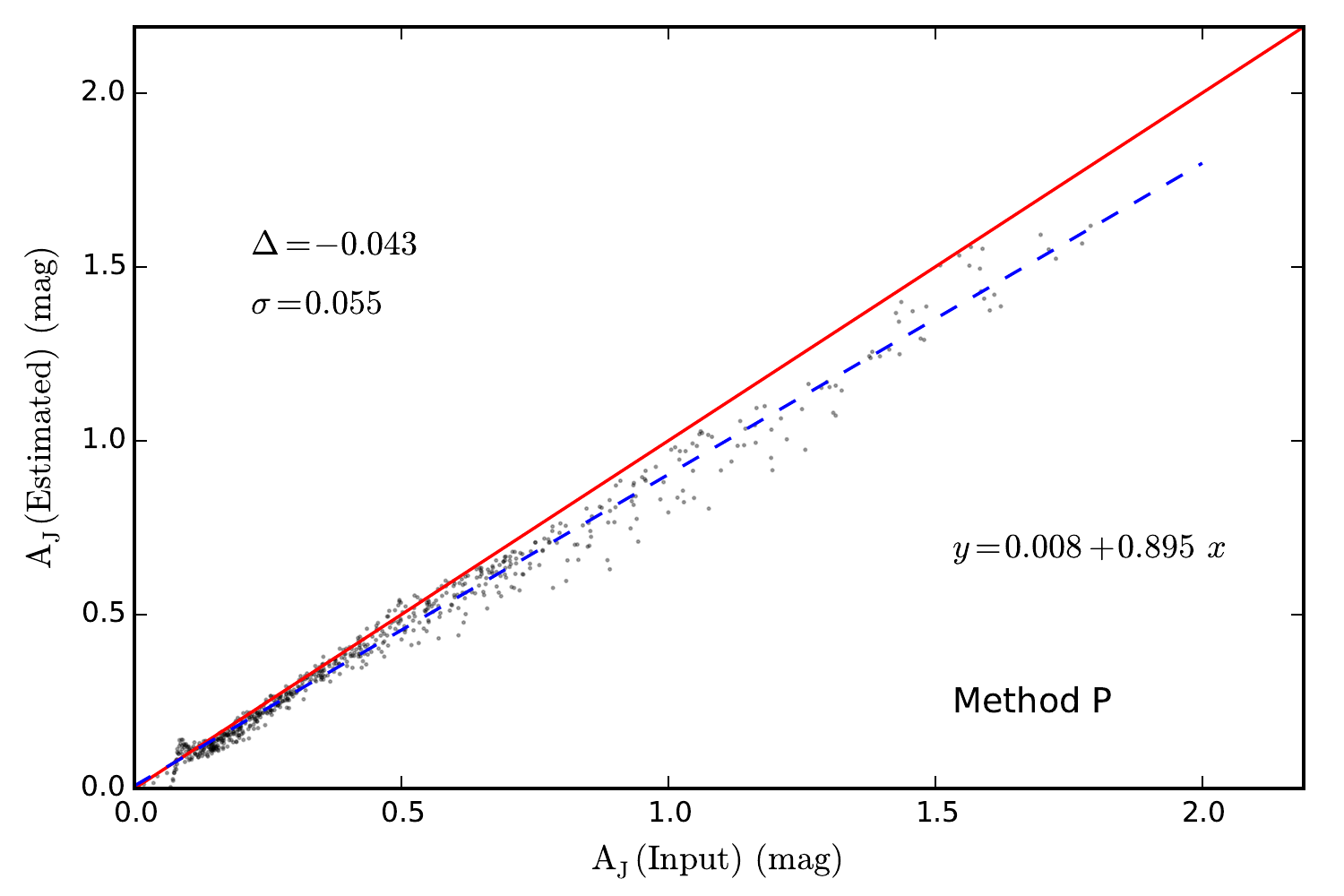} 
\caption{
Result of Method P, using {\em \emph{a priori}} classification into two source populations.
}
\label{fig:P2}
\end{figure}

Figure~\ref{fig:P2} shows that the explicit treatment of the two populations has
reduced both the rms error and bias but not by much. Compared to Method B in
Fig.~\ref{fig:1B_map_sca}, the overall bias has decreased from -0.050\,mag to -0.043\,mag,
and the rms error has decreased from 0.064\,mag to 0.055\,mag. 
In the present case, each beam is likely to contain several sources from both
populations so that when Method B fits a single $A_J$ value, the result is already
unique. If the solution assigned sources of one category to a wrong peak in $P_C$, the
sources of the other category would fall completely outside the distributions, resulting
in a very low overall probability. Therefore, the advantage of Method P should be
greater if the populations overlap in colour-colour space so that there is a greater
risk of confusion or if the number of sources per beam is so small that, without a
priori information of their category, they could all be attributed to either of the two
source populations.

\subsection{Tests using Besan\c{c}on simulations} \label{sect:1MMX}

\begin{figure*}
\includegraphics[width=17.0cm]{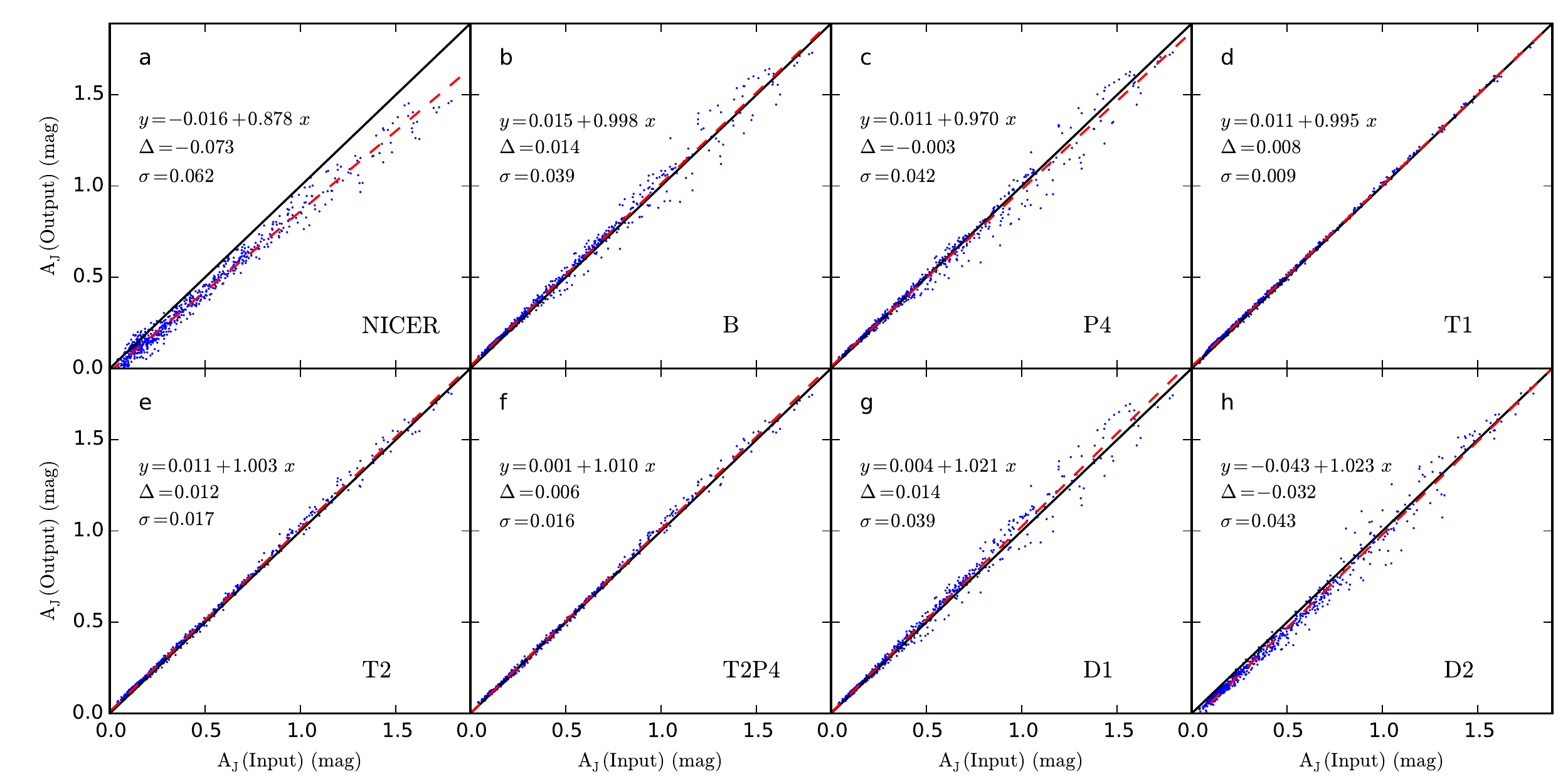} 
\caption{
Correlations between estimated and true extinction in tests with stars simulated with
the Besan\c{c}on model. Results are shown for NICER (frame a), Method B (frame b),
Method B with separate reference colours for four magnitude intervals (frame C), Method
T using a perfect extinction template (frame d), or using the NICER map as the template with
a single (frame e) or four reference colours (frame f). The final frames include bias
correction with methods D1 and D2 (frames g and h, respectively). Each frame lists the
parameters of linear least squares, the bias $\Delta$, and the rms error $\sigma$.
}
\label{fig:BE}
\end{figure*}

The final test uses a more realistic case where the intrinsic colours are obtained from
the Besan\c{c}on stellar population model \citep{Robin2014}. The synthetic catalogue only
consists of stars, but their colour distribution depends on Galactic location and
apparent magnitude. We used the data as shown in Fig.~\ref{fig:plot_BES}, which
corresponds to coordinates ($l$, $b$)=(10$\degr$, 20$\degr$). As illustrated by the
figure, the colour-colour distribution is elongated and changes significantly as a
function of apparent magnitude. If the photometric accuracy of the observations is very
high, this information could be used in calculations to define probability distributions
separately for different magnitude intervals. To emphasise the potential differences
between the methods, we simulated very deep observations with low photometric errors. The
completeness limit is set around 23 magnitudes in the $H$ band, and the photometric errors
are 0.1 times the typical 2MASS errors ($k_{\rm noise}=0.1$). Such observations may be
impossible for current ground-based instruments but may become possible in the near future
either with
ELT\footnote{http://www.eso.org/sci/facilities/eelt/instrumentation/index.html} or 
space-borne
instruments\footnote{http://people.lam.fr/burgarella.denis/denis/\\2014\_WISH\_+\_First\_Galaxies\_Workshop.html}.
Results of the tests conducted in this section are summarised in Fig.~\ref{fig:BE}.

\begin{figure}
\includegraphics[width=8.8cm]{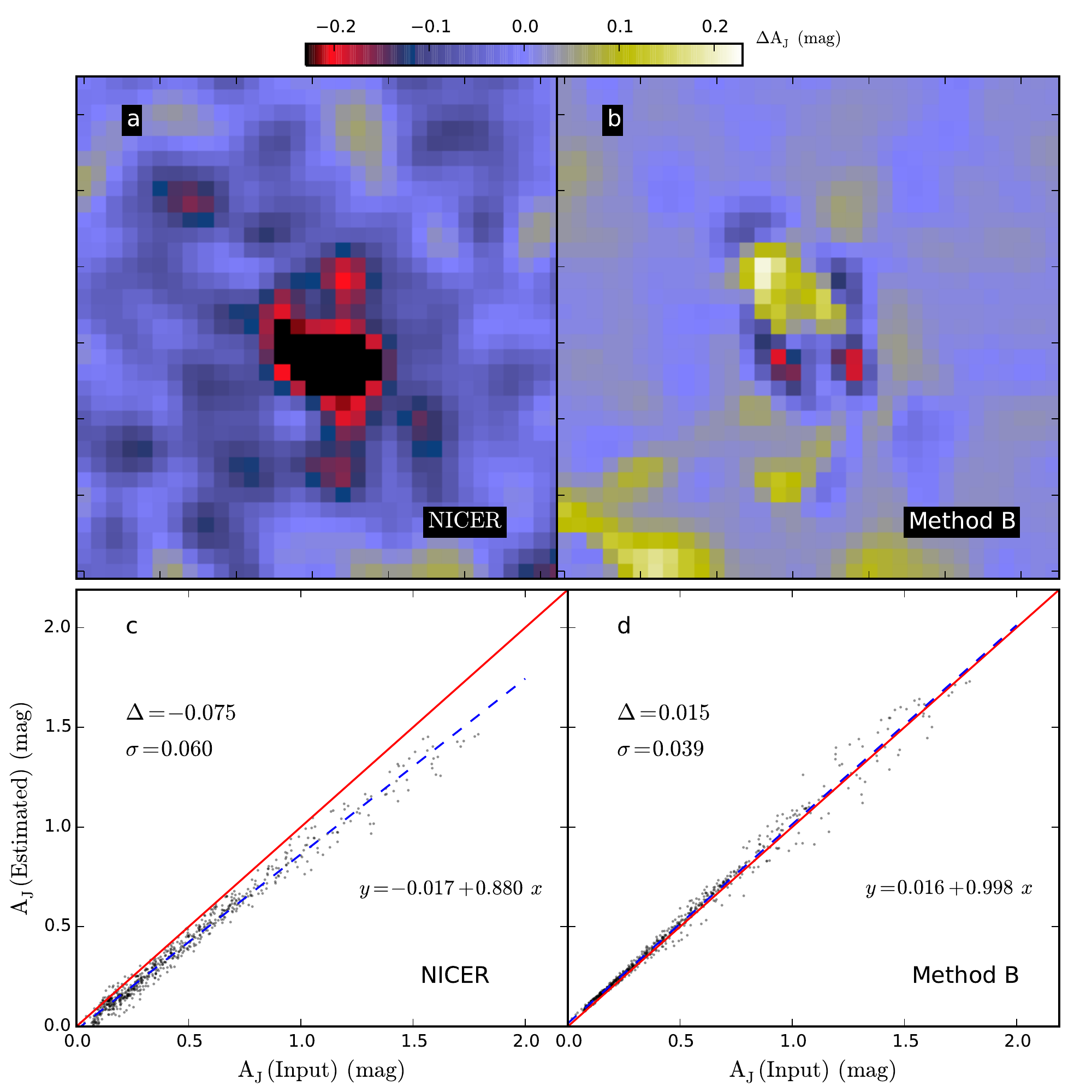} 
\caption{
Comparison of NICER (left) and Method B (right) results in the case of stars simulated
with the Besan\c{c}on model. The upper panels show maps of absolute error $\Delta A_J$ and
the lower panels the correlations between recovered and true $A_J$ values.
}
\label{fig:PB2}
\end{figure}

Figure~\ref{fig:PB2} shows the results for NICER and Method B. For the realisation
used in these tests, the reference NICER solution has a bias of -0.08\,mag and an rms
error of 0.06\,mag. Method B results in a bias of 0.02\,mag and an rms error of
0.04\,mag. The rms value of Method B is smaller mainly because of the good performance
below $A_J\sim$0.5\,mag. The absolute accuracy of Method B estimates is
also better at
high extinctions, the comparison with the input extinction map giving a linear slope of
$\sim1.0$. For both methods, the largest errors are concentrated near the extinction
maximum. As already discussed, this is caused by column density gradients and results
from the false model assumption of a constant extinction across the beam. 

Although not evident in these tests (or in Sect~\ref{sect:T2}), Method B could encounter
some problems when intrinsic colours are concentrated along a very narrow ridge in the
colour-colour plane. If probability drops very fast outside the ridge, a single star
(observed through an extinction that is slightly different from the beam-average value)
can have a strong influence on the estimate of the beam-averaged extinction. Methods
like Method T should be more robust because they explicitly include sub-beam extinction
fluctuations as part of the model, and a crude fix would be to convolve $P_C$ to
accommodate some small-scale variations in $A_J$. However, as shown by
Fig.~\ref{fig:plot_BES}, in practice this was not a significant problem.

Tests in Sect.~\ref{sect:1T} show that the small-scale column density variations
are a significant source of noise in the extinction maps. If the relative $A_J$ values
within the beam are known, Method T drastically reduces the errors (see
Fig.~\ref{fig:BE}d). For the same realisation as above, the rms noise is reduced by a
factor of four and the estimates remain accurate up to the largest column densities.
The average bias is 0.008\,mag, smaller than for Method B, which already was quite close
to zero. Because we use here the true extinction map as an input, it is not something
that could be applied to real observations, at least not without reliable,
high-resolution ancillary data (see Sect.~\ref{sect:method_T}).
If we use the NICER extinction map as a template of the small scale structure (mainly
gradients), the improvement of rms noise is still quite significant, a factor of two
over Method B (Fig.~\ref{fig:BE}e). These estimates remain accurate even near the
central peak.

We also tested Method P, using separate $P_C$ distributions for stars in the apparent
magnitude intervals $<15$, 15-19, 19-22, $>$22\,mag. The limits are selected so that a
roughly equal number of stars falls within each category (apparent magnitudes without
extinction). With the subdivision Method P results in bias $\Delta=$-0.003\,mag and rms
error $\sigma=$0.042\,mag, with a least squares slope of 0.97 (Fig.~\ref{fig:BE}c). In
other words, the result is slightly worse than with Method B and a single $P_C$
distribution. This may be because, once the reference stars are divided to four samples,
each individual $P_C$ map has somewhat higher noise.
Using the same four categories with Method T and NICER template results in a bias of
$\Delta=0.006$\,mag and an rms noise of $\sigma=$0.016\,mag (Fig.~\ref{fig:BE}f), close
to the result obtained using Method T and a single $P_C$ probability distribution. Thus,
in the case of these more realistic colour-colour distributions, there does not appear
to be any advantage in using multiple $P_C$ distributions. 

This may be surprising considering the large variation seen in Fig.~\ref{fig:plot_BES} and the degeneracy of
solution for any star located either in the horizontal or in the vertical branch of that
figure. In our test most stars are in the horizontal part, but each beam also contains
many stars so that the value of the beam-averaged extinction is not ambiguous. The
presence of stars in both branches always defines a unique solution. If the stellar
density were very low, the {\em \emph{a priori}} information provided by the apparent
magnitude could be expected to become more important.

Methods D1 and D2 try to reduce bias without explicit information of sub-beam column
density structures. In Fig.~\ref{fig:BE} Method B results were already almost unbiased
and neither D1 nor D2 results in clear improvement. The least squares slope between
estimated and true extinction is 1.02 for both D1 and D2. The scatter $\sigma$ is
similar to Methods B and thus larger than for Method T.

\subsection{Summary of results on simulated observations} \label{sect:simusum}

\begin{figure}
\includegraphics[width=8.8cm]{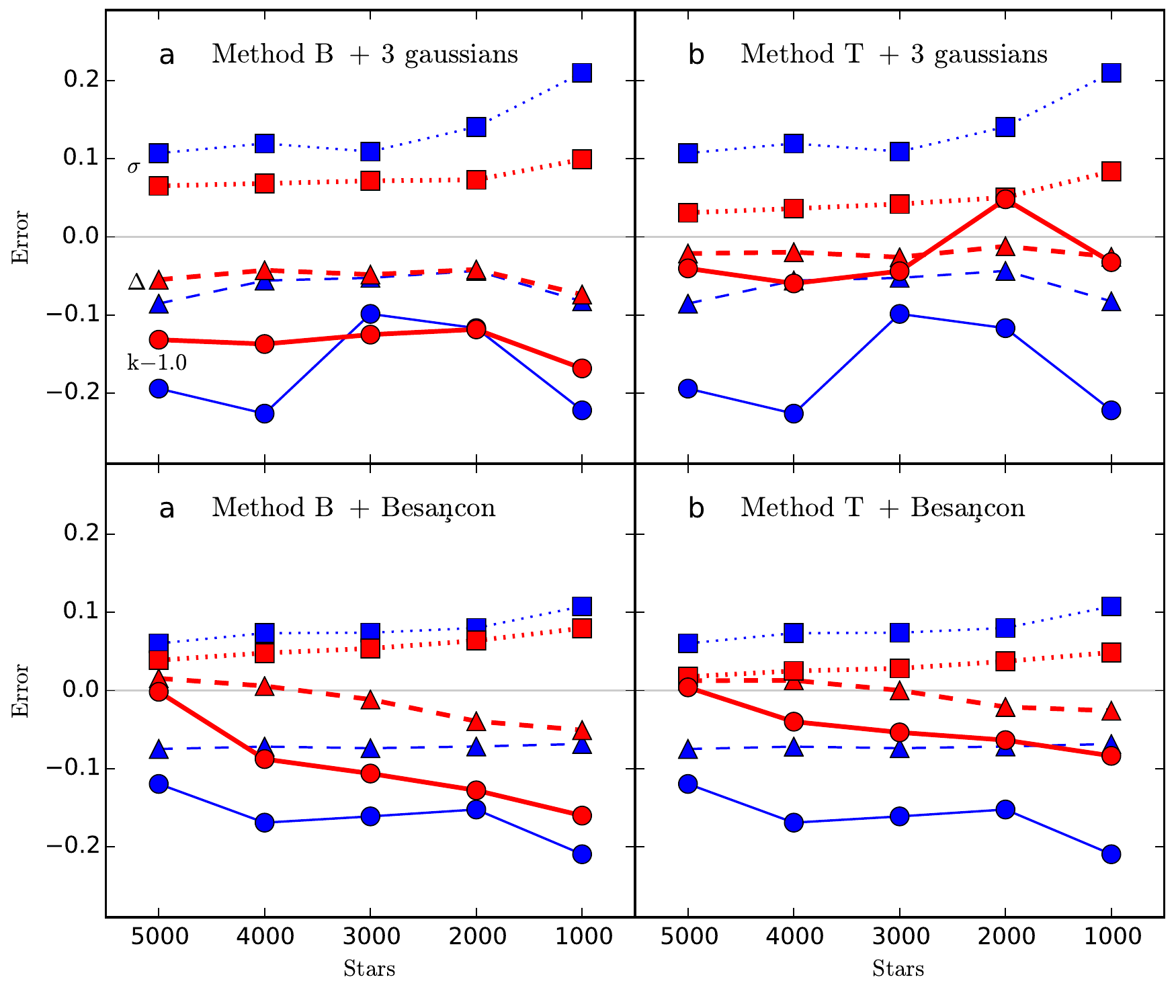} 
\caption{
Comparison of rms error $\sigma$ (squares), bias $\Delta$ (triangles), and least squares
slopes $k$ (circles; error is $k$-1.0) for NICER (blue symbols and lines) and SCEX (red
symbols and thick lines) as a function of stellar density. Each decrease of 1000 stars
per map also corresponds to an increase of 1.0\,mag in the limiting magnitude. 
}
\label{fig:plot_series}
\end{figure}

The previous tests show that if the distribution of intrinsic colours is not
approximated well with a Gaussian distribution, it is useful to use a discretised
representation to retain full information of the distribution.
Figure~\ref{fig:plot_series} summarises the results for the colour distributions
discussed above, one consisting of three Gaussian components and the other one based on
stars simulated from the Besan\c{c}on model. Figure~\ref{fig:plot_series} shows a series
of calculations with decreasing stellar density. In the previous examples the number of
stars was 5000 over the map area. In Fig.~\ref{fig:plot_series} we decrease the number
of stars in steps of 1000 down to 1000 stars per map. At the same time we change the
detection threshold so that each step corresponds to a 1.0\,mag increase in the value of
the limiting magnitude (removal of the faintest stars).  The simulations with 2MASS
stars start with the original completeness curves of the survey. In the original
distribution of Besan\c{c}on stars, the faintest stars are $\sim$24.5\,mag in $H$ band,
but because of the completeness curves applied in the simulations, the peak in the
counts is for more than 0.5\,mag brighter stars.
Because the intrinsic colours depend on the apparent magnitude, a change in the
detection threshold also means a change in the average colour (cf.
Fig.~\ref{fig:plot_BES}).

Figure~\ref{fig:plot_series} shows that the rms error (measuring difference of the
recovered extinction map and the true input map) is always lower for Method B than for
NICER. The difference is greater in the case of colour-colour distributions consisting of
three Gaussians, i.e., strong deviation from a distribution that could be approximated
with a single covariance matrix. Compared to Method B, Method T again performs
consistently better. The improvement is most noticeable in the bias (average difference
between estimated and true extinction) and in the slope of the least square fits that
compares extinction estimates with the known true values. The change from 5000 to 1000
stars per map does not cause a very strong increase in errors, and in most cases the
increase shows a similar rate for both NICER and Methods B and T. One exception is
seen for simulated Besan\c{c}on stars where the Method B (and to some extent Method T)
estimates start with a much lower bias but approach the NICER bias values as the number
of stars is decreased. It is important to note that a decrease in stellar density is associated
here with the removal of the faintest stars that have a colour distribution
that is very different from the brighter stars. Thus, the remaining stars have more
similar colours.

\subsection{Application to real observations} \label{sect:example}

As a final test we examine extinction maps of the Pipe Nebula. The NICER maps of the
region are available already on line\footnote{Lombardi, Marco; Alves, Joao; Lada,
Charles J., 2014, ``2MASS extinction map of the Pipe Nebula'',
http://dx.doi.org/10.7910/DVN/25112 Harvard Dataverse Network V3}, since the calculations
are based on stars from the 2MASS survey \citep[see][]{Lombardi2006_Pipe}. These maps
have a resolution of 1.0$\arcmin$ and a pixel size of 30$\arcsec$. We compare
calculations where the distribution of intrinsic colours is derived either from 2MASS
stars in a reference region or from the Besan\c{c}on model. 

We start by using reference colours estimated with $\sim$46~000 2MASS stars in the
vicinity of Galactic coordinates (355.97, 8.00). Because of the large photometric errors
(compared to those used in Sect.~\ref{sect:1MMX}) the reference colours can be
approximated with a 2D Gaussian, and results of Method B are expected to be similar to
those of NICER. For the same reason, the use of several $P_C$ distributions is not
likely to be relevant. Therefore, we concentrate on Method T.

\begin{figure*}
\includegraphics[width=18.5cm]{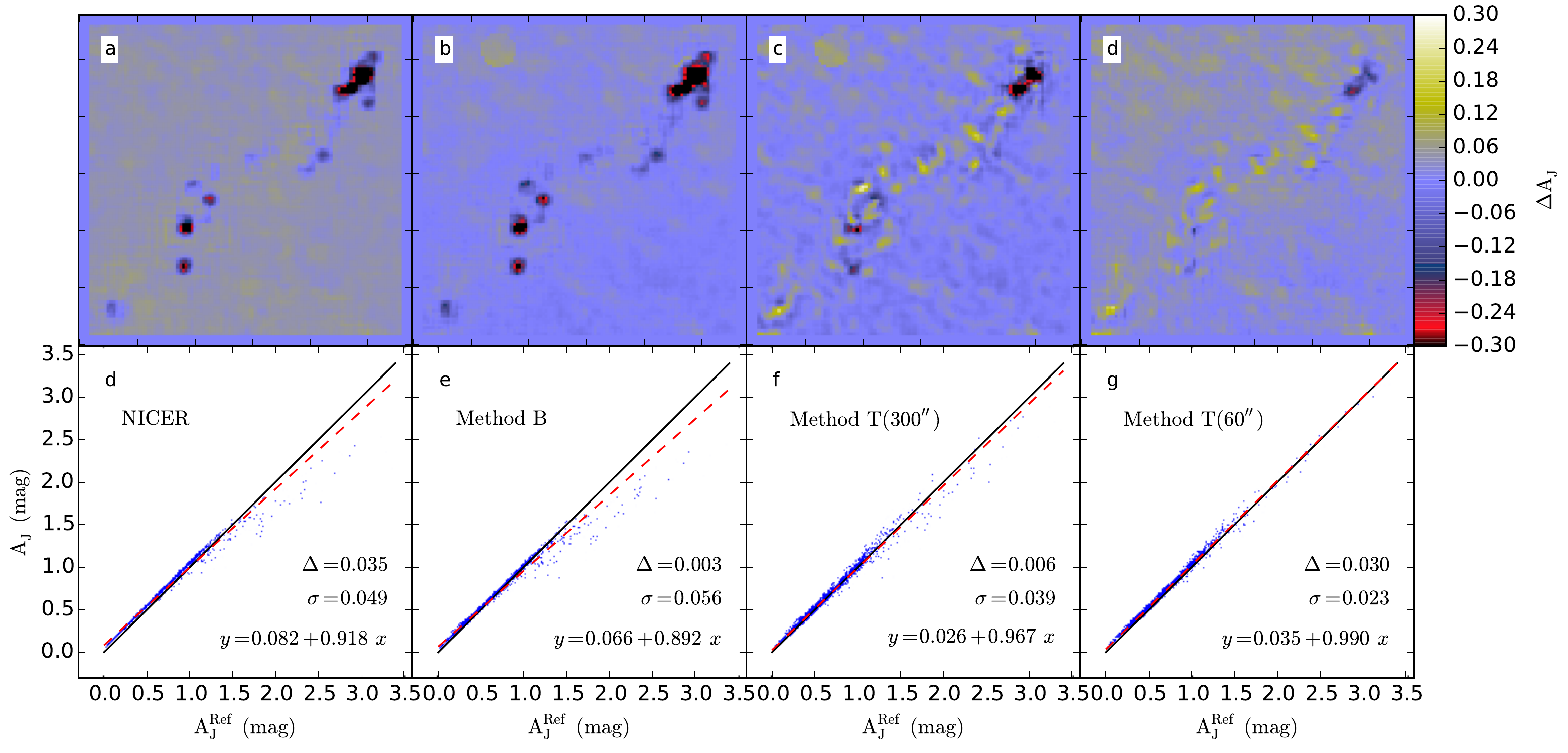} 
\caption{
Residuals for extinction maps calculated for the Pipe Nebula at a resolution of 
300$\arcsec$. The reference is the \citet{Lombardi2006_Pipe} NICER map that has
been convolved to the resolution of 300$\arcsec$ (see text). Residuals are shown for
NICER (frame a), Method B (frame b), and Method T (frames c and d). Method T uses the NICER map of frame a (frame c)
as a template of the small-scale structure or the
\citet{Lombardi2006_Pipe} NICER map at the full 60$\arcsec$ resolution (frame d).
Frames d-g show the corresponding data as scatter plots with the reference data on the
x-axis. The maps are in Galactic projection and have a size of 3.22$\times$3.22 degrees.
}
\label{fig:PIPE}
\end{figure*}

Figure~\ref{fig:PIPE} shows the results for extinction maps calculated at a resolution
of 300$\arcsec$. The large beam size was selected because in this case, the results are
expected to show significant bias and underestimate the extinction of the dense clumps.
The results are compared to the NICER map originally calculated at 60$\arcsec$
resolution and then convolved to the 300$\arcsec$ resolution. As expected, NICER
and Method B give rather similar results, although the slope of the linear fit is 0.03
units lower for Method B, and this suggests that $P_C$ can in this case be approximated
with a single Gaussian. Both methods underestimate the extinction of the reference map,
the error at the highest extinction reaching $\sim$30\% when compared to the reference,
which is the \citet{Lombardi2006_Pipe} map convolved down to the 300$\arcsec$
resolution. Method T (frames c and f) shows improvement over NICER and Method B. The rms
error is lower by $\sim30\%$, and especially the bias of the highest extinction values
has decreased. In frames c and f, the template was the NICER map that was convolved to a
resolution of 300$\arcsec$. The last frames of Fig.~\ref{fig:PIPE} show the results when
the template is the reference map with a 60$\arcsec$ resolution. In this case the slope
of the linear fit is very close to unity.

\begin{figure*} 
\includegraphics[width=18.5cm]{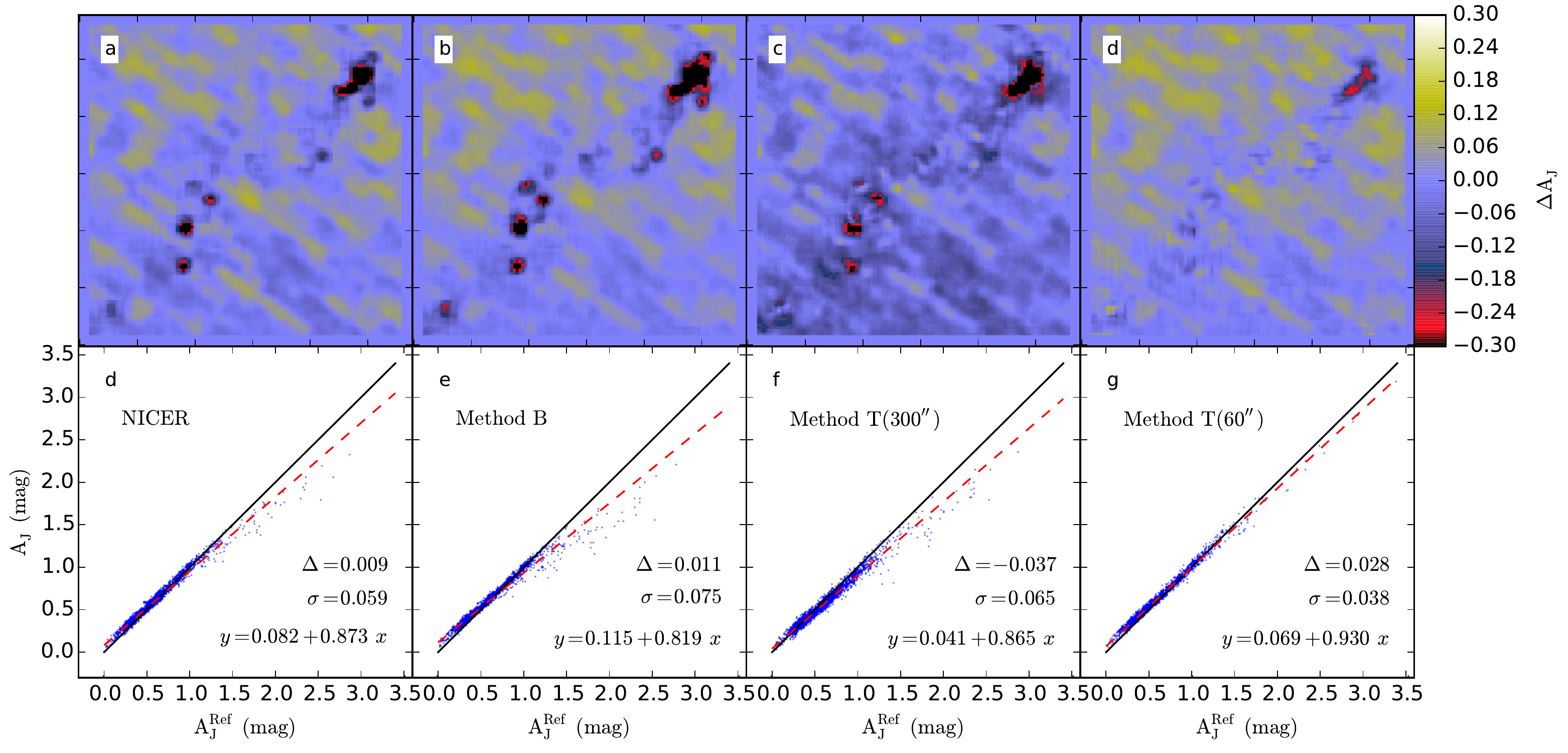}
\caption{
As in Fig.~\ref{fig:PIPE} but using distributions of reference colours based on the Besan\c{c}on
model.
}
\label{fig:PIPE_BE}
\end{figure*}

We repeat the calculation by extracting $P_C$ from stellar catalogues simulated with
the Besan\c{c}on model for the centre of the field. In Fig.~\ref{fig:PIPE_BE} the
reference (x-axis of the scatter plots) is again the \citet{Lombardi2006_Pipe} NICER map
convolved to 300$\arcsec$ resolution. In Fig.~\ref{fig:PIPE_BE}a our NICER map uses the
same simulated reference stars as the SCEX calculations, which accounts for the 
difference between frames d of Figs.~\ref{fig:PIPE} and \ref{fig:PIPE_BE}.

In Fig.~\ref{fig:PIPE_BE} the absolute extinction values are lower than in
Fig.~\ref{fig:PIPE}.  
For the first time, Method B appears to perform worse than NICER. We believe this to be
a sign of problems in the description of reference colours. Based on simulated 
stars, $P_C$ is constructed
directly, and the distribution of intrinsic colours is narrower
than the colours of the actual stars in the field (almost a factor of two
difference when measured along the H-K axis). This is also thus true in a direction
perpendicular to the reddening vector. For stars outside the expected colour-colour
distribution, the calculated probabilities will correspond to the tail of the $P_C$
probability distribution, i.e., the tail of the Gaussian kernel that was used to convert simulated
stars into a continuous $P_C$ distribution (see Sect.~\ref{sect:methods}). The
probability gradients along the reddening vector are therefore largely arbitrary for
those stars whose intrinsic colours are outside the assumed colour-colour distribution.
The situation is somewhat different for NICER, where the distribution of intrinsic colours is
described with a single 2D Gaussian, which is also affected by the general shape of the
colour-colour distribution and not only by the local probability profile along a given
reddening vector. In the present case the mismatch between $P_C$ and the distribution of
observed stars is fairly obvious. Nevertheless, the result suggests that NICER may be
generally more robust against this type of inaccuracy.

In Fig.~\ref{fig:PIPE_BE}, even Method T does not show any improvement over NICER when
the 300$\arcsec$ resolution NICER map is employed as a template. Only when the template
with 60$\arcsec$ resolution is used are the results finally better in terms of both
the scatter $\sigma$ and the slope of the least squares line.
The improvement is most significant for dense clumps.

In the case of Fig.~\ref{fig:PIPE_BE}, one must remember that the reference (x-axis of
the scatter plots) is a NICER calculation where the intrinsic colours are derived from
observations of a real reference field. Therefore, some of the differences are
caused by the different assumptions of the intrinsic colours.

\begin{figure}
\includegraphics[width=8.8cm]{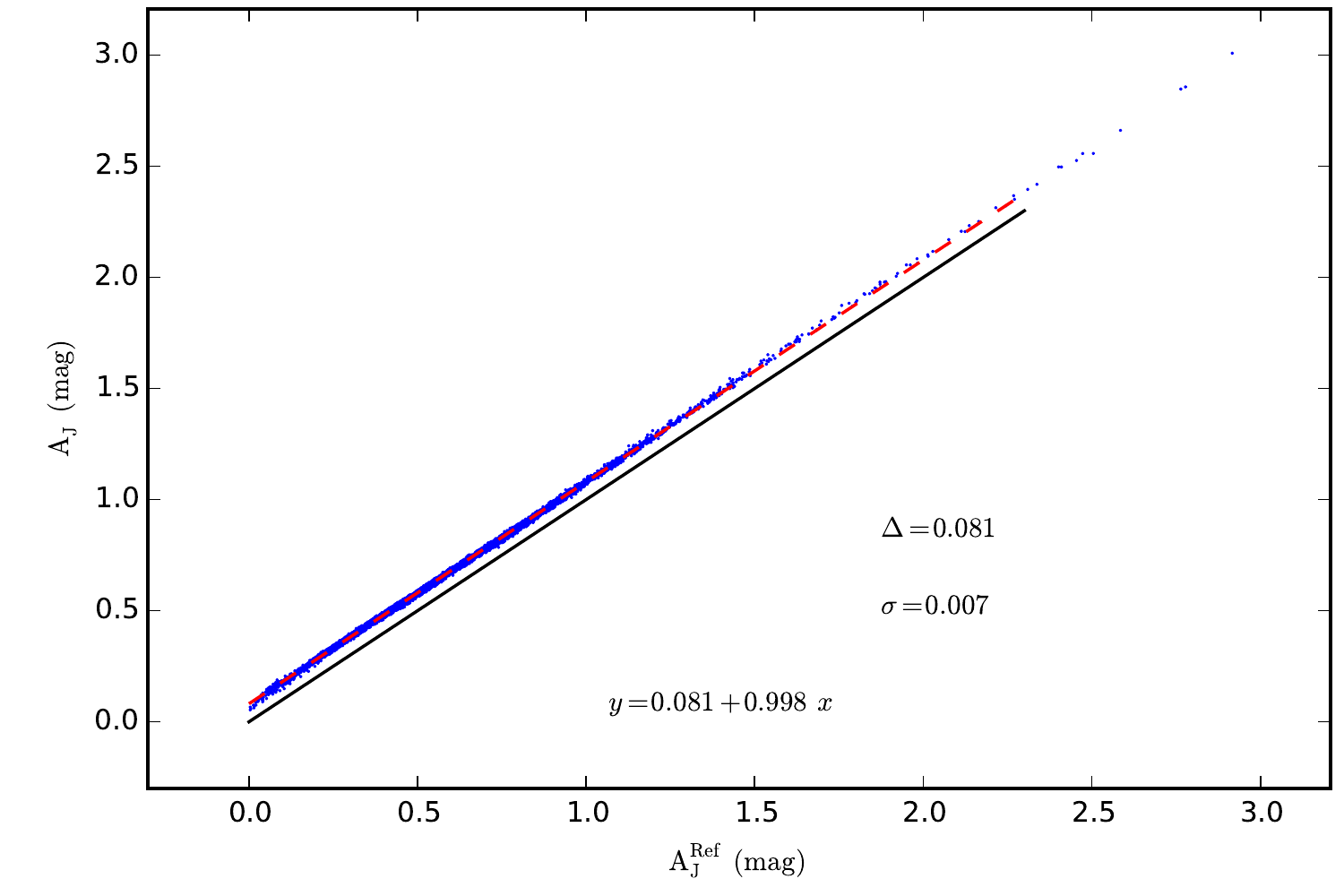}
\caption{
Comparison of Method B with reference colours based on the Besan\c{c}on model. The colours
correspond to two locations along the filament, one below the central latitude (x-axis
of the figure, reference position $l=$358.8, $b=$5.6) and one near the northern end
(y-axis of the plot, reference position $l=$357.1, $b=$7.1).
}
\label{fig:PIPE_NS}
\end{figure}

Figure~\ref{fig:PIPE_NS} shows the correlation between extinction estimates calculated
with Method B with reference colours extracted from the Besan\c{c}on model at two 
reference positions. The positions correspond to the main clumps within the map area and
are separated by a distance of 2.2 degrees (1.5 degrees in Galactic latitude).  The
figure shows that the gradient of intrinsic colours results in a zero-point shift that
is less than $\Delta A_J=0.1$\,mag over the whole field. Nevertheless, if the
effect is neglected, this causes an observable gradient over the extinction map. The
effect is $\ga$10\% for all values below $A_J=1.0$\,mag, causing the calculated
extinction to drop slightly too fast as a function of Galactic latitude. Second, the rms
scatter between the two calculated extinction maps is 0.7\%. This is caused in part by
noise in the reference colour distributions (derived from a finite sample of simulated
stars) but shows an upper limit for the Monte Carlo noise in our calculations.

\section{Discussion}  \label{sect:discussion}

We have examined extinction calculations with the observed NIR colours of background
sources. The starting point was the premise that instead of using a single covariance
matrix to describe the distribution of intrinsic colours of the sources, it may be
better to operate with discretised probability distributions. We studied the use of
three near-infrared bands, using observed $J-H$ and $H-K$ colours and their expected
distribution in a 2D colour-colour plane.
Within the MCMC framework we have tested several modifications that might be useful. In
particular, these include (1) use of {\em \emph{prior}} information on small-scale column
density structure, (2) debiasing using a model for the statistics of column density
variations or using a NICEST type correction, and (3) division of sources into
subcategories, for example, based on apparent magnitudes.

The tests confirmed that when the distribution of intrinsic colours is approximately
Gaussian, the methods recover the same result as NICER calculations (e.g.,
Fig.~\ref{fig:T1_00_scat}). Some differences should arise already if the distribution of
intrinsic colours is skewed, and this also depends on our decision to use the median value
of MCMC samples as the extinction estimate. The tests in Sect.~\ref{sect:T2} were chosen
to create conditions where the difference to the simple Gaussian approximation is very
pronounced. If a large number of the sources were galaxies that cannot be filtered out
prior to the extinction analysis, the situation would be rather similar to that of
Fig.~\ref{fig:1B_info}. The first conclusion of these tests is that, in spite of the
wrong assumption (regarding the distribution of intrinsic colours), NICER results remain
quite accurate. Nevertheless, Method B shows noticeably lower rms noise, especially at
low extinctions. The quantitative results also depend on the structure of the column
density distribution. Both methods underestimate the extinction in regions of strong
extinction gradients. 
If the photometric errors are increased or the intrinsic colour distribution is
more Gaussian, the differences in the noise at low column densities disappear. This is also
the case for the actual observations shown in
Figs.~\ref{fig:PIPE}-\ref{fig:PIPE_BE}. In tests like the one shown in
Fig.~\ref{fig:1B_map_sca}, the noise of NICER and SCEX maps is also similar at higher
column densities where the errors appear to be dominated by the random sampling by
background stars rather than the uncertainty of the intrinsic colours.

In the presence of column density gradients, the estimates of Method B are biased
downwards, and the errors are similar to those of NICER maps. One of the most promising
improvements of Method B is the use of ancillary information on sub-resolution variations
in extinction. Method T uses the ratio of extinction along a given LOS (a single star)
and the average extinction over a beam. Therefore, the estimated extinction values
remain independent of the scaling of the ancillary data, and only estimates of relative
extinction values are needed. Furthermore, the ratios are calculated independently for
each beam, again avoiding any direct link between the large-scale morphology of the
reference data and the estimated extinction map. 

In Method T, the explicit inclusion of information on the small scale was seen to also be
very effective in decreasing the bias. Although perfect knowledge of sub-beam
structure can be expected to improve the accuracy of extinction estimates, its impact is
still surprisingly large (see Fig.~\ref{fig:T1_T2}). This is particularly clear when the
photometric errors are small and most of the noise results from small-scale column
density variations. When the extinction map itself is used as the reference, the
correction is still usually effective, especially if the column density variations are
mostly caused by large-scale gradients that are also resolved by low-resolution
extinction data. In Fig.~\ref{fig:T1_T2}, at low column densities, the errors are a
fraction of the corresponding errors in NICER. When the reference data fail to resolve
important column density peaks, the errors are still smaller than for NICER, but the
scatter is higher, and the estimates are biased towards lower values. The only case where
Method T did not produce clear improvements is shown in Fig.~\ref{fig:PIPE_BE}f.
However, when the template map had a higher resolution, Method T also performed well in
this test.

High-resolution estimates of column density can be available from various sources, such
as molecular line or dust continuum observations. For example, Herschel observations
cover most of the nearby molecular clouds, providing dust column density maps down to a
resolution of 18$\arcsec$ resolution or even better. If the resolution of extinction
maps is at most $\sim 1\,\arcmin$, this should be enough to provide significant
improvement in the accuracy of extinction estimates. On the other hand, in the case of
deep NIR observations, the resolution of extinction maps themselves may be so high that
no higher resolution reference data exists. However, even the use of extinction data
itself can sometimes result in a significant improvement in accuracy (e.g.,
Fig.~\ref{fig:T1_T2}b).

The changes in the distribution of intrinsic colours (e.g., Fig~\ref{fig:plot_BES})
have suggested that extinction estimates could be improved by considering the
differences in the intrinsic colours of stars of different apparent magnitudes. In
practice no clear improvement was observed. We suspect that the main cause is that, as
long as there are many stars within the beam, the solution is well constrained even
without this additional information. For an individual star, there may be more than one
possible solution along the reddening vector. For example, in Fig.~\ref{fig:plot_BES}
an unextincted star might reside either in the horizontal part (low extinction) or
in the vertical part (much higher extinction) of $P_{\rm C}$. If the beam contains a
single star, Method B returns the median of the probability distribution, which in this
case contains two peaks. Thus, the estimate is likely to be either clearly too low or
clearly too high (similar to NICER if one approximates the intrinsic colour distribution
with a single Gaussian). However, as soon as the beam contains stars originating in both
the vertical and the horizontal parts of the distribution, the solution of the
beam-averaged extinction becomes unique. Furthermore, the stars in different magnitude
intervals do not have entirely different intrinsic colours. They simply fill in parts of
an already well-constrained common probability distribution. Thus, the situation is
similar to Sect.~\ref{sect:T2} where it is not 
necessary to know beforehand which
population an individual source belongs to. The situation could change if the number of
sources per beam is very small so that all sources could by chance belong to one of the
two populations.

We also tested Method D1 where the probability distribution between the beam-averaged
extinction and the extinction seen by an individual star was explicitly taken as a part
of the model. In Method D1, the asymmetry of this distribution was established using the
cumulative star numbers and an assumption of the column density statistics. The method
should reduce the bias caused by sub-beam column density variations. The main weakness
of this method is that the magnitude of the correction directly depends on the
assumed magnitude of the column density fluctuations. In the practical tests, Method D1
did not result in significant improvements. On the other hand, Method D2 implements the
principles of the NICEST method \citep{Lombardi2009_NICEST}. The correction only depends
on the extinction estimates calculated for individual stars, i.e., on observable
quantities. In tests, Method D2 resulted in some improvement (especially in 
Fig.~\ref{fig:D1_D2_scat}b), but it can also lead to increased rms errors (see
Fig~\ref{fig:BE}h).

\begin{figure}
\includegraphics[width=8.8cm]{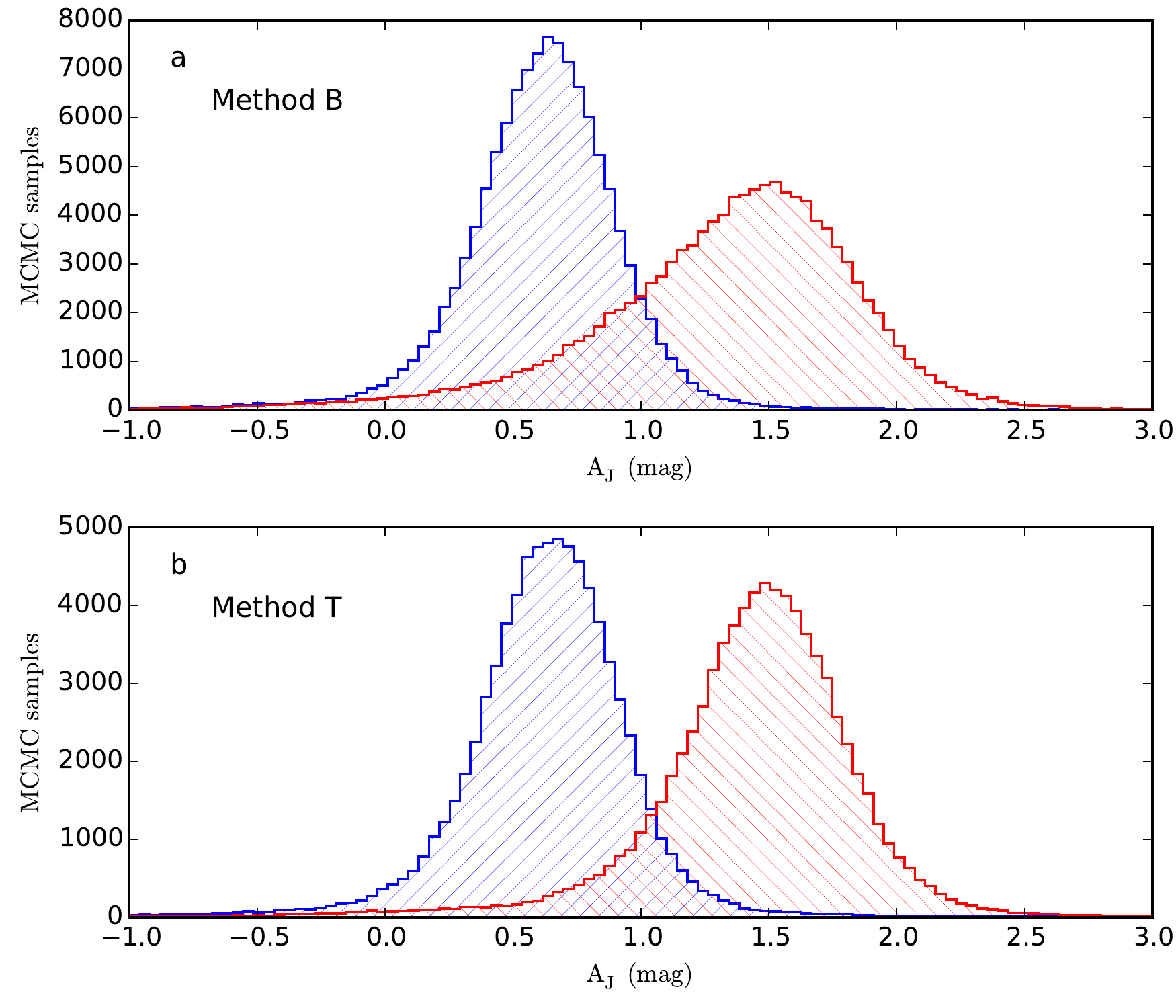}
\caption{
Probability distributions for the extinction in two individual pixels (blue and red
histograms) in the maps discussed in Sect.~\ref{sect:T2}. Distributions are shown
for Method B (upper frame) and Method T (lower frame) for the same pixels.
}
\label{fig:prob}
\end{figure}

MCMC has the advantage of providing estimates of the full posterior probability
distribution of extinction. On the other hand, the computations are slower, in some
cases by orders of magnitude, compared to NICER or NICEST methods. However, all methods
discussed in this paper could also be used in connection with faster least squares or
optimisation methods (e.g., on top of existing NICER or NICEST implementations; see
\ref{appendix:allsky}). This would reduce, although not eliminate, the difference in
computational time but would also reduce the information we have of the parameter
distributions. In this context, it is also interesting to examine the actual posterior
probability distributions of the extinction values. Figure~\ref{fig:prob} shows examples
for Methods B and T. 
Because of the way the photometric errors were included in calculations (see
Sect.~\ref{sect:methods}), the ratios of calculated probabilities are not strictly
correct. This affects the width of the posterior probability distribution,
therefore Fig.~\ref{fig:prob} should be taken only as a qualitative indication of the
differences in these distributions.
At low extinction values, which in this case also correspond to a relatively constant
extinction across the beam, the distributions are almost Gaussian. Larger deviations
from normal distribution are seen for the high extinction (and high $A_J$ gradient)
pixels. As indicated by the rms values, Method T results in smaller uncertainties. This
is also associated with a more Gaussian probability distribution as the long tail to
lower extinction values is decreased. This suggests that especially with Method T,
faster calculations (least squares instead of MCMC) would not necessarily result in a
significant loss of information.

\section{Conclusions}  \label{sect:conclusions}

We have examined the calculation of extinction based on the colours of background stars.
We characterised the colours of unextincted stars using a full 2D probability
distribution, instead of the conventional Gaussian approximation. We examined several
variations of the basic method.
The study led to the following conclusions:
\begin{itemize}
\item By replacing Gaussian approximation with a more accurate description of intrinsic
source colours, one can significantly reduce the noise of the extinction maps. If the
sources only consist of stars, photometric errors can eliminate most of this advantage. 
However, a precise description of the intrinsic colours will be important in case of
high-precision measurements from future instruments such as ELT or WISH.
\item 
The largest improvements are obtained by including {\em \emph{a priori}} information about the
small-scale column density structure (Method T). This could be in the form of higher
resolution observations of dust emission. However, by tracing the large-scale gradients,
the information contained in the extinction data itself is sometimes sufficient to
significantly reduce both the bias and the noise of the extinction estimates.
\item
Method D2 can be very efficient in correcting the bias of extinction maps. This has
already been demonstrated before because, apart from the use of a discretised intrinsic
colour distributions, the method is similar to NICEST. However, if a good template map
is available (one with low noise and preferably high resolution), Method T can result in
an equally low bias and smaller rms errors.
\end{itemize}
We carried out the computations using the Markov chain Monte Carlo program SCEX. However,
the ideas can be used just as well in connection with faster least squares or optimisation
methods, making the methods more practical for analysing large areas.

\begin{acknowledgements}
This publication makes use of data products from the Two Micron All Sky Survey, which is a
joint project of the University of Massachusetts and the Infrared Processing and Analysis
Center/California Institute of Technology, funded by the National Aeronautics and Space
Administration and the National Science Foundation.
MJ acknowledges the support of the Academy of Finland Grants No. 250741 and 285769.
\end{acknowledgements}

\bibliography{biblio_v1.3}

\appendix

\section{All-sky NICER map employing Method T} \label{appendix:allsky}

As an example of implementing some of the discussed ideas outside the MCMC framework, we
have expanded the previous calculation of an all-sky NICER map
\citep{JuvelaMontillaud2015a} to employ Method T. A normal NICER map was first
calculated on Healpix pixelisation \citep{Gorski2005} with a resolution of 
FWHM=3.0$\arcmin$ and with resolution parameter NSIDE=4096, which corresponds to a pixel
size of $\sim$0.86$\arcmin$. This map was then used as a template in a second
calculation. When the $A_J^i$ values of individual stars were combined to calculate the
estimate of the beam-averaged extinction $A(J)$, the values $A_J^i$ were first scaled by
the ratio $A(J)/A_J^i$ read from the template map. As discussed in
Sects.~\ref{sect:method_T} and~\ref{sect:1T}, this is most useful for correcting 
large-scale gradients that are resolved by the template map. The previous extinction map
cannot of course probe actual structure on scales below the beam size, and the correction
is further limited by the pixel size, which is only a factor of three smaller than the
FWHM of the final map. 

\begin{figure}
\includegraphics[width=8.8cm]{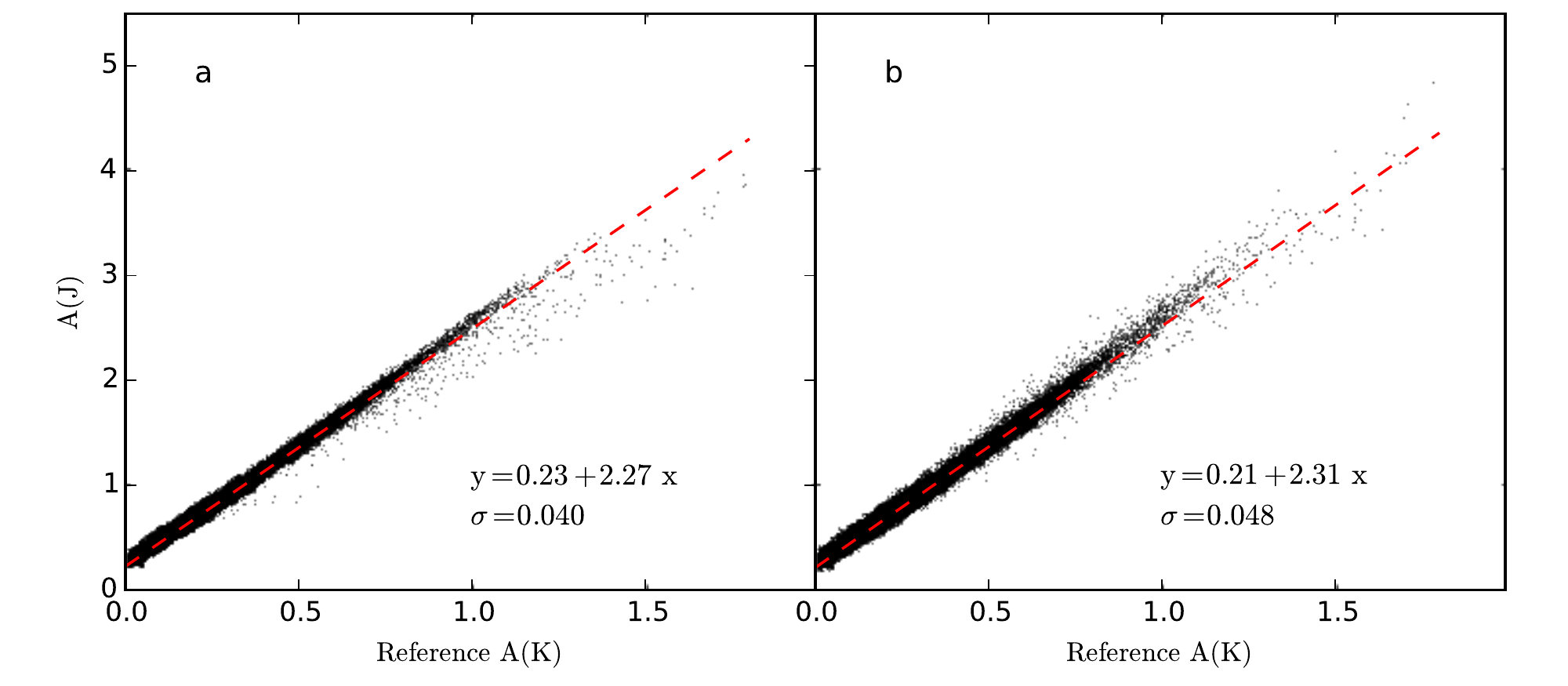}
\caption{
Comparison of our all-sky extinction maps $A(J)$ and the \citep{Lombardi2006_Pipe} NICER
map given as $K$-band extinction. The latter has been convolved from the original
1.0$\arcmin$ resolution down to 3.0$\arcmin$ resolution. The all-sky map has been
calculated with the NICER method (frame a) or with a modified method using Method T
(frame b).
}
\label{fig:cmp_template}
\end{figure}

Figure~\ref{fig:cmp_template} compares the first NICER map and the second, Method T, map
of the Pipe Nebula region. Both data are plotted against values from the
\citep{Lombardi2006_Pipe} NICER map of the Pipe Nebula, which was originally calculated
at a resolution of 1.0$\arcmin$ and was for this comparison convolved down to a
resolution of 3.0$\arcmin$. As discussed in \citet{JuvelaMontillaud2015a}, this should
result in much lower bias than when a NICER map is calculated directly at a low
resolution. In the latter case, which is applicable to our all-aky maps, the average extinction
will be underestimated because the number of detected stars systematically decreases
with column density. This is visible in Fig.~\ref{fig:cmp_template}a, where at high
column densities our values fall below the least squares line fitted to all data points.
When Method T is used (frame b), the scatter of the relation increases but is more
symmetric with respect to the least squares line, which is also somewhat steeper than in
frame a. The relation remains linear up to the highest extinction values where the
values are up to $\sim$20\% higher than with the basic NICER method. This suggests that
Method T is able to improve the extinction estimates using only the photometry data,
without any external information. When implemented in connection with NICER, the impact
on computation times is minimal, and this enables calculation of large extinction
maps\footnote{The all-sky extinction map is available in electronic form at the CDS via
anonymous ftp to cdsarc.u-strasbg.fr (130.79.128.5) or via
http://cdsweb.u-strasbg.fr/cgi-bin/qcat?J/A+A/. It can also be found at
http://www.interstellarmedium.org/Extinction.}.

%
%

\end{document}